\documentclass[onecolumn,notitlepage,english,prl,showpacs,superscriptaddress,longbibliography]{revtex4-1}
\usepackage[utf8]{inputenc}
\usepackage{placeins}
\usepackage[T1]{fontenc}
\usepackage{lmodern,amsmath,amssymb,amsfonts,bm,graphicx,color,mathrsfs}
\usepackage[colorlinks,citecolor=blue,linkcolor=blue,urlcolor=blue]{hyperref}
\usepackage[mathlines]{lineno}
\newcommand{\Tr}{\operatorname{Tr}}
\newcommand{\cN}{\mathcal N}
\setcitestyle{super,sort&compress}
\hypersetup{pdftitle={Quantum Recovery from Black-Hole Radiation Beyond the Emission Yield},pdfauthor={Kashif Ammar Yasir}}
\begin{document}
\title{Quantum Recovery from Black-Hole Radiation Beyond the Emission Yield}
\author{Kashif Ammar Yasir}
\email{kayasir@zjnu.edu.cn}
\affiliation{Department of Physics, Zhejiang Normal University, Jinhua 321004, China}
\affiliation{Zhejiang Institute of Photoelectronics, Jinhua 321004, China}
\date{28 October 2026}
\begin{abstract}
A proposed resolution of the black-hole information paradox must explain which information an outside observer can recover from escaping radiation\cite{Hawking,Page,Hayden,Almheiri}. Environmental memory can return correlations to the source without delivering them to that observer\cite{ChenQi,ChenMemory}. Here we establish a sharp recovery frontier at fixed mean radiation yield for single-excitation channels and realize its qubit limit with an autonomous quantum-memory emitter. Unequal transmissions balance retained coherence against information in failed collection events. A reference--radiation correlation test distinguishes this gain even when the mean collected energy is unchanged. Identical transmission channels instead give the same recovery at matched collection, regardless of their emission timescales. For a broader class of interacting sources, an ordering of collected and inaccessible coupling spectra forbids unassisted quantum transmission irrespective of internal memory. A separate coding example supports a positive rate where uniform transmission at matched actual yield has zero capacity. Together, these results identify what openness must change to assist recovery: the accessible transmission structure. They supply a physically constrained comparison of evaporation mechanisms, separating information delivered to radiation from coherence retained elsewhere or an increase in the collected output.
\end{abstract}
\maketitle

\section*{Information recovery from an open source}
Can quantum memory improve what an observer recovers from radiation, rather than merely return coherence to the emitting source? This question matters for black-hole information and for quantum transmission through lossy environments. Hawking spectra and fine-grained entropy characterize important aspects of evaporation\cite{Hawking,Page,Almheiri}. A receiver additionally needs a channel that preserves an unknown state's entanglement with a reference. Correlations retained by an inaccessible environment do not automatically contribute to that channel. A claim that quantum openness helps resolve the paradox therefore needs a comparison at fixed observer access and collected output, together with a signal that distinguishes the resulting channels.

Typical subsystem states can look thermal while a global state remains pure\cite{PageAverage}; gravitational entanglement-wedge calculations likewise reconcile entropy with information-preserving evaporation in controlled settings\cite{Penington,AlmheiriQES}. These results motivate a sharper operational comparison: hold the collected output budget fixed and optimize recovery using that output alone. Exact and approximate correction connect this task to information in the complement\cite{KnillLaflamme,Kretschmann,Beny}, whose role is also exposed by the no-hiding theorem\cite{NoHiding}. A source-coherence revival, an entropy turnover or a reversal involving the environment answers a different access question.

A microscopic open-system description derives fluctuations and response from common environmental degrees of freedom\cite{FeynmanVernon,HuPazZhang} and retains the emitted field through input--output theory\cite{GardinerCollett}. Quantum Langevin treatments of cavity condensates and magnomechanics illustrate the corresponding link between dissipative coupling, response and entanglement\cite{YasirCavity,YuMagnon}. Here we retain the field's state and correlations to calculate the receiver's channel, rather than infer information recovery from source relaxation.

Stochastic gravity and noise-driven horizon or atmosphere models constrain geometric and field observables\cite{HuVerdaguer,YasirUnruh,YasirAtmosphere}. Those tests are complementary to the phase-sensitive source-to-radiation map studied here. Our result concerns how much quantum information reaches a specified observer, with the uncollected output included in the same microscopic description.

\begin{figure}[!htbp]
\centering\includegraphics[width=\textwidth]{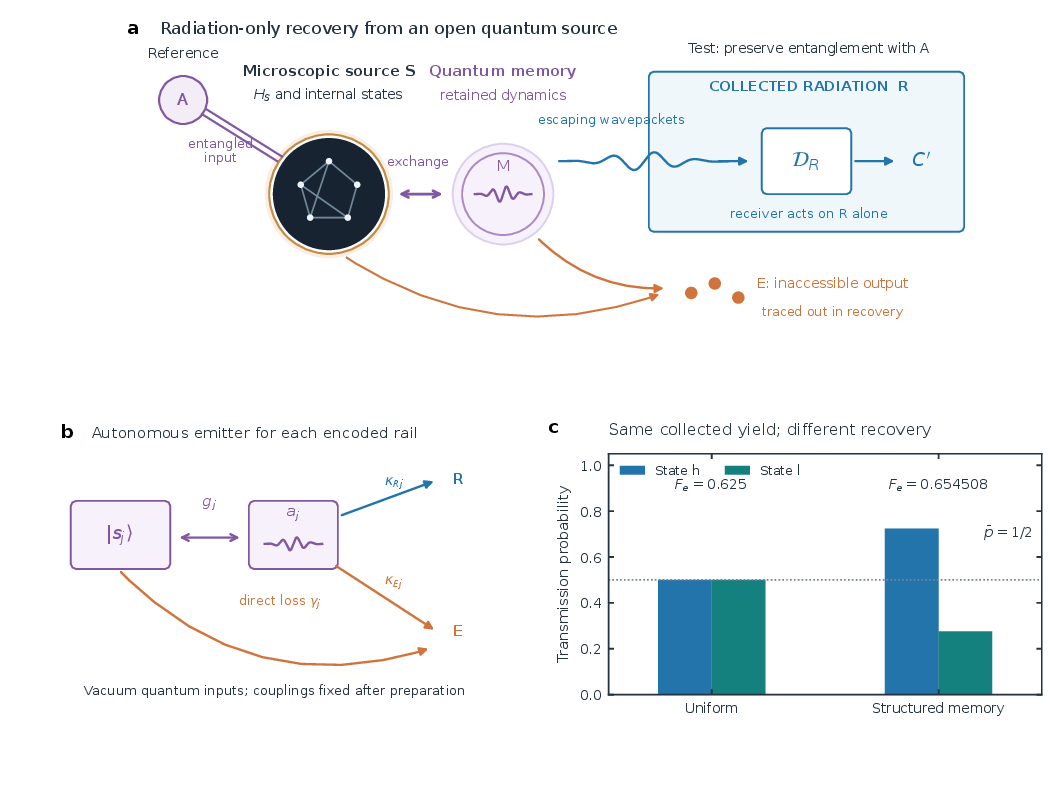}
\caption{\textbf{Quantum memory, restricted access and the recovery task.} \textbf{a}, The reference $A$ is initially entangled with information encoded in source $S$. Retained quantum memory $M$ exchanges excitations with the source. An observer receives escaping radiation $R$ and applies a decoder $\mathcal D_R$, aiming to preserve the input's entanglement with $A$. Terminal inaccessible output $E$ is traced out. The dark source disk is a schematic icon for internal quantum states, rather than a calculated spacetime image. \textbf{b}, The autonomous one-excitation realization: source state $|s_j\rangle$ couples to memory oscillator $a_j$ with strength $g_j$; memory decay partitions into $R$ and $E$, and direct source loss enters $E$. Inputs are vacuum and couplings stay fixed after preparation. \textbf{c}, Equal mean collection, $\bar p=1/2$, permits different transmission spectra and therefore different deterministic recovery fidelities. The structured example attains the exact qubit frontier.}\label{fig:overview}
\end{figure}

We consider an observer who processes the complete designated outgoing field $R$, while the source remainder and an inaccessible output $E$ are traced out (Fig.~\ref{fig:overview}). The source evolves autonomously after preparation. This access restriction differs from environment-assisted correction\cite{Gregoratti} or deliberate source--reservoir reversal. A code of dimension $k$ is initially maximally entangled with a reference $A$. Its deterministic recovery fidelity is
\begin{equation}
 F_\star(\cN)=\max_{\mathcal D_R}
 \langle\Phi_k|(\operatorname{id}_A\otimes\mathcal D_R\circ\cN)(\Phi_k)|\Phi_k\rangle,
 \qquad |\Phi_k\rangle=k^{-1/2}\sum_{j=1}^k|j\rangle_A|j\rangle_C .
 \label{eq:fidelity}
\end{equation}
The decoder is completely positive and trace preserving, so failed collection events remain in the result. This is Schumacher entanglement fidelity optimized over recovery\cite{Schumacher}, a single-use quantity distinct from the asymptotic unassisted quantum capacity $Q$.

The reference tests coherence without giving the decoder knowledge of the input. Both arrival and vacuum outcomes count, and the receiver may process the complete designated radiation jointly. The source remainder, retained memory and terminal environment are unavailable. This gives an upper benchmark for any receiver with additional bandwidth or detection constraints.

Retained internal degrees of freedom give quantum channels a memory\cite{KretschmannMemory}. Measures of backflow and operational Markov conditions characterize related aspects of quantum processes\cite{Breuer,Pollock}, but need not order them by radiation recovery. A longer-lived reservoir can also hold an excitation near a lossy source and reduce collection. The useful comparison therefore fixes the input and records the collected yield explicitly.

Decoding makes information retrieval operational once dynamics and receiver resources are specified\cite{YoshidaKitaev}. Bosonic-loss theory\cite{Caruso,MultiCaruso,BradlerCapacity}, generalized erasure\cite{Filippov}, resonant quantum communication\cite{Catalano}, and microscopic source--bath models\cite{ChenQi,ChenMemory} supply the foundations. Our central result is a sharp fixed-yield recovery frontier with autonomous qubit attainment. A separate matrix condition excludes unassisted transmission for arbitrary interacting sources within the stated bath model. Together with exact subspace optimization, these results separate a useful transmission imbalance from a change in collection time or total yield.

\section*{A microscopic recovery obstruction}
Let $H_S$ describe the source, $\bm O$ its coupling operators and $\bm b_\alpha(\omega)$ independent bosonic input fields. In units $\hbar=1$,
\begin{align}
 H&=H_S+\sum_{\alpha=R,E}\int_0^\infty d\omega\,
 \omega\bm b_\alpha^\dagger\bm b_\alpha+H_I,\nonumber\\
 H_I&=\sum_{\alpha=R,E}\int_0^\infty d\omega\,
 [\bm b_\alpha^\dagger G_\alpha(\omega)\bm O+\mathrm{h.c.}],
 \qquad J_\alpha=G_\alpha^\dagger G_\alpha .
 \label{eq:microscopic}
\end{align}
The regulated Hamiltonian is assumed well defined, with initially independent vacuum inputs and asymptotic outgoing channels. Integrating the bath operators gives both quantum noise and emitted amplitudes (Methods). Thus response, fluctuations and the radiation state originate in the same interaction.

The matrices $J_R$ and $J_E$ specify how each source operator couples to the two output groups. Their eigenvectors need not coincide, so comparing scalar decay rates is generally insufficient. A generalized eigenvector of $(J_R,J_R+J_E)$ identifies a combination of source couplings whose emitted field is split with a definite collection fraction. Directions with zero total spectral density are uncoupled and are removed before taking the inverse square root. This construction permits multiple coupling operators, noncommuting spectral matrices and an interacting source, while retaining independent vacuum inputs.

The bath rotation is a change of field coordinates, not an action that the observer applies to the source. It groups the amplitudes generated by the same source operator into a coupled field and an orthogonal vacuum field. An observer's restriction to $R$ then appears as loss after otherwise arbitrary coupled-output dynamics. This order of operations is the reason a channel statement can be proved without solving the source's many-body time evolution.

Set $J=J_R+J_E$. The eigenvalues of $J^{-1/2}J_RJ^{-1/2}$ on the support of $J$ are collection transmissions $\eta_j(\omega)$. A frequency-preserving rotation separates coupled bath combinations from unused vacuum modes. The source-to-radiation channel then factors into the complete coupled output followed by bosonic attenuation with these transmissions. All source interactions and memory reside in the first map. For transmission at most one-half, attenuation is antidegradable: the inaccessible output can simulate the collected output\cite{Caruso,MultiCaruso}. Consequently,
\begin{equation}
 \boxed{J_R(\omega)\preceq J_E(\omega)\quad\text{at every coupled frequency}
 \quad\Longrightarrow\quad Q(\cN_R)=0,\qquad F_\star\le\frac{k+1}{2k}.}
 \label{eq:nogo}
\end{equation}
The fidelity ceiling follows from the same impossibility of unrestricted quantum copying that underlies optimal cloning bounds\cite{WernerCloning}. Antidegradability supplies the channel structure needed to apply this restriction\cite{Cubitt}. It holds for every allowed isometric code and deterministic receiver decoder, including codes in an interacting many-body source. Additional detector loss preserves the conclusion.

\begin{figure}[!htbp]
\centering\includegraphics[width=\textwidth]{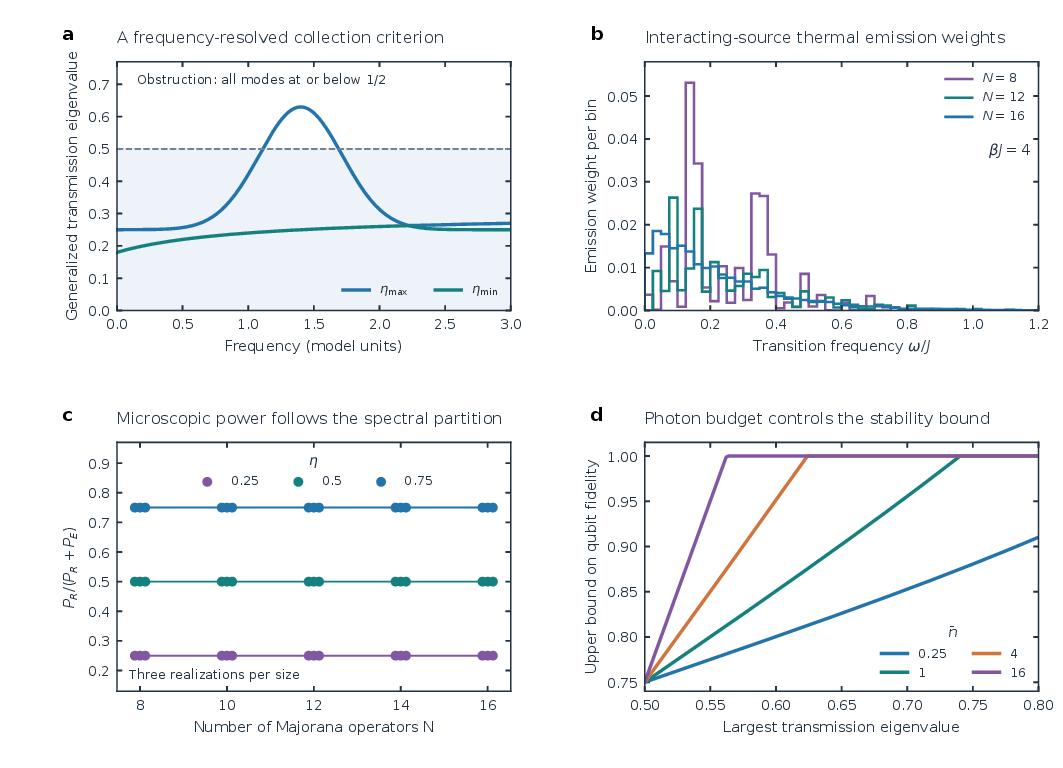}
\caption{\textbf{Microscopic spectral tests and a robust recovery ceiling.} \textbf{a}, Generalized collection eigenvalues for the noncommuting spectral-matrix example specified in Supplementary Information, S11. The shaded region has transmission at most one-half; Eq.~\eqref{eq:nogo} applies when all coupled modes lie in that region. The above-threshold window removes that sufficient obstruction. \textbf{b}, Binned exact thermal emission weights of finite SYK sources, at $\beta J=4$. All transitions are retained in the source data. \textbf{c}, Golden-rule power partition for proportional bath spectra, with three disorder realizations per size. Points are slightly offset horizontally for visibility. The lines are exact algebraic predictions, independent of microscopic source weights. \textbf{d}, The conservative fidelity ceiling for a qubit, as a function of the largest transmission eigenvalue and coupled-output photon budget $\bar n$. The bound is capped by unity.}\label{fig:spectral}
\end{figure}

Equation~\eqref{eq:nogo} translates the established loss threshold into a sufficient condition on microscopic coupling matrices. Changing a common spectral envelope, and hence its memory time, cannot remove this obstruction. A violation opens a possibility for recovery; it does not itself prove a positive rate. Small violations admit a photon-budget-dependent fidelity bound (Methods and Fig.~\ref{fig:spectral}d). Reflected fields that return to the source belong to retained dynamics, whereas $E$ denotes terminal loss. This distinction is essential when applying the criterion to gravitational scattering and greybody transmission.

Physically, antidegradability says that the discarded output is at least sufficient to construct the receiver's state. If that receiver could also reconstruct an arbitrary quantum input perfectly, the two complementary outputs would supply two copies of its entanglement with the reference. The impossibility of such copying supplies the obstruction. Zero unassisted capacity still allows nonzero finite-use fidelity and classical communication. It also does not cover protocols that acquire the inaccessible environment or additional assisting resources. The conclusion therefore identifies a precise boundary for openness-based mechanisms: memory must change the effective partition or transmission structure, rather than merely reshuffle amplitudes behind the same ordered loss map.

A source with a controlled gravitational interpretation is the quartic Sachdev--Ye--Kitaev (SYK) Hamiltonian,
\begin{equation}
 H_{\rm SYK}=\sum_{i<j<l<m}J_{ijlm}\chi_i\chi_j\chi_l\chi_m,
 \qquad \{\chi_i,\chi_j\}=\delta_{ij},\qquad
 \overline{J_{ijlm}^2}=6J^2/N^3 .
 \label{eq:syk}
\end{equation}
At low energy and large $N$, its reparametrization mode is governed by a Schwarzian action, with a connection to nearly AdS$_2$ gravity\cite{Maldacena,Kitaev,MaldacenaYang}. We couple a parity-preserving bilinear to a weak super-Ohmic bath; conformal power counting makes this specified probe infrared irrelevant (Supplementary Information, S11). Exact spectra at $N=8$--$16$ test source operators, detailed balance and initial emission weights. For proportional spectral matrices, the computed power partition reproduces the exact collection ratio for every sampled source (Fig.~\ref{fig:spectral}). The channel obstruction itself is independent of source size and of the weak-coupling approximation used for this power diagnostic.

Figure~\ref{fig:spectral} separates three levels of evidence. Its frequency-resolved example shows how a matrix coupling can contain an above-threshold window even when other modes are strongly lost. The SYK line spectra show the source dependence of the initial emission weights. Their proportional-bath power ratios provide an algebraic check that those microscopic weights cancel in the predicted partition. This cancellation is a consistency result, not evidence for a universal Page curve. The finite spectra and the exactly solved emitter answer different parts of the argument: the former instantiate the source coupling, while the latter determine a complete, phase-sensitive output map.

\section*{The fixed-yield recovery frontier}
A single excitation emitted into orthogonal radiation modes supplies an exactly resolved benchmark. Let $T$ map the source code into the collected one-particle space. Keeping complete temporal wavepackets and tracing inaccessible systems yields the generalized erasure channel\cite{Filippov},
\begin{equation}
 \cN_T(\rho)=T\rho T^\dagger+
 \Tr[(I-T^\dagger T)\rho]|\mathrm{vac}\rangle\langle\mathrm{vac}|,
 \qquad A_T=T^\dagger T\le I .
 \label{eq:erasure}
\end{equation}
The vacuum event is orthogonal to successful collection. Their separate optimization gives
\begin{equation}
 \boxed{F_\star(T)=\frac{(\Tr\sqrt{A_T})^2+1-\lambda_{\min}(A_T)}{k^2}.}
 \label{eq:exactF}
\end{equation}
The receiver undoes the isometric part of $T$ on the successful output and prepares a minimum-transmission eigenstate on vacuum. Failure contains partial classical information about which state was sent, because weakly transmitted states fail more often. Exploiting that information raises deterministic average recovery without assuming that a lost coherent state has been reconstructed.

The optimum balances two competing effects. Unequal transmission distorts the collected coherent state, reducing the squared-trace contribution in Eq.~\eqref{eq:exactF}. It also makes failure informative about the encoded basis state, increasing the best deterministic fallback contribution. The frontier solves that competition at fixed mean yield. Its gain is therefore a property of the complete decoded channel, including failure, rather than of a selected set of successful events.

\begin{figure}[!htbp]
\centering\includegraphics[width=\textwidth]{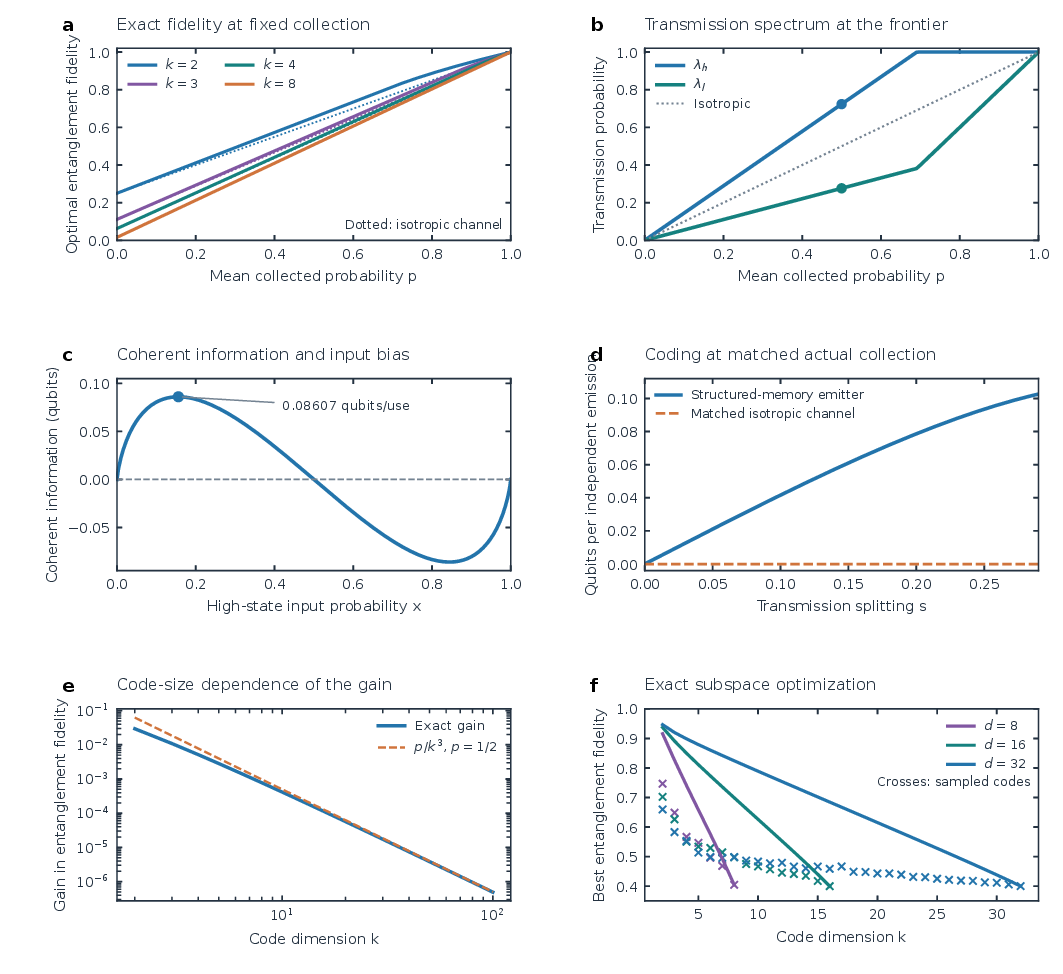}
\caption{\textbf{Sharp deterministic recovery and a separate coding benchmark.} \textbf{a}, Exact entanglement-fidelity frontiers for the indicated code dimensions; dotted curves show equal transmission at the same mean collection. Failed events are included. \textbf{b}, Optimal qubit transmission eigenvalues, with markers at half collection. \textbf{c}, Coherent information for that spectrum versus diagonal-input bias. Its positive maximum gives an achievable rate of $0.08607388$ qubits per independent preparation; this is not the exact quantum capacity. \textbf{d}, Rate lower bounds for spectra $(1/2+s,1/2-s)$. Each isotropic comparator is matched to the actual collected yield of the corresponding biased input and has zero unassisted capacity. \textbf{e}, Exact fidelity advantage at $p=1/2$ and its $p/k^3$ large-code asymptote. \textbf{f}, Optimal isometric subspace codes for the test spectra $\lambda_j=[0.05+0.90(j-1)/(d-1)]^{1.2}$. Lines are the analytical optima; crosses show the best of 150 sampled complex isometries for each dimension pair. The sampled points check the implementation, while the optimization theorem establishes the bound over every permitted subspace. Panels~a,b,e,f concern single-use entanglement fidelity; panels~c,d concern coding across independent preparations. All comparisons retain the designated radiation and exclude the inaccessible environment.}\label{fig:frontier}
\end{figure}

The encoding optimization is also exact. If the full allowed source space has ordered transmission eigenvalues $\lambda_1^\downarrow\ge\cdots\ge\lambda_d^\downarrow$, the optimal $k$-dimensional isometric code uses its top $k$ eigenvectors. Its fidelity is Eq.~\eqref{eq:exactF} with those eigenvalues. Eigenvalue interlacing and coordinatewise monotonicity prove optimality over all such subspace codes; sampled encodings only check the implementation (Fig.~\ref{fig:frontier}f). Superselection restrictions are imposed before this optimization.

At fixed mean collection $p=\Tr A_T/k$, the sharp frontier has $k-1$ equal high transmissions $b$ and one low transmission $m$, where
\begin{align}
 t_k&=\tfrac12\left(1-\sqrt{(k-1)/(k+3)}\right),\quad
 m=\max\{kp\,t_k,kp-(k-1)\},\quad b=\frac{kp-m}{k-1},\nonumber\\
 F_{\max}(k,p)&=\frac{[(k-1)\sqrt b+\sqrt m]^2+1-m}{k^2}.
 \label{eq:frontier}
\end{align}
Concavity first makes the transmissions above the minimum equal; optimizing that minimum completes the proof. For a qubit at half collection,
\begin{equation}
 (\lambda_h,\lambda_l)=\tfrac12(1+1/\sqrt5,1-1/\sqrt5),\qquad
 F_{\max}=\frac{3+\sqrt5}{8}=0.6545084972 .
 \label{eq:qubit}
\end{equation}
Equal transmission gives the ordinary erasure channel\cite{BennettErasure}, with $F_{\rm iso}=p+(1-p)/k^2=0.625$. The absolute gain, $0.0295085$, balances the information carried by failure against distortion of the successful coherent state (Fig.~\ref{fig:frontier}a,b). For fixed $p<1$, it decreases as $p/k^3$ at large code dimension (Fig.~\ref{fig:frontier}e). The optimization concerns entanglement fidelity; its input-state trade-off is visible in Fig.~\ref{fig:dynamics}d.

This is a sharp benchmark for the stated single-excitation channel class. The proof gives both an upper bound and a spectrum that attains it, rather than an improvement relative to a selected decoder. Its gain decreases with code dimension; it does not imply a constant fractional advantage for an arbitrarily large initial state. Figure~\ref{fig:frontier}e,f shows the complementary encoding result: within a larger allowed source space, its strongest transmission eigenvectors form the optimal small subspace code.

\section*{Autonomous memory and matched emission}
The solvable emitter contains excited source states $|s_j\rangle$, a common ground $|g\rangle$ and memory oscillators $a_j$. Fixed couplings $g_j$ exchange an excitation between $|s_j\rangle$ and $a_j$. Each oscillator leaks into collected and inaccessible quantum continua at rates $\kappa_{Rj}$ and $\kappa_{Ej}$, while source decay at rate $\gamma_j$ enters $E$ (Fig.~\ref{fig:overview}b). Eliminating the retained oscillator produces the exponential quantum-memory kernel of a pseudomode model\cite{Garraway,Pleasance}. Quantum input commutators and output wavepackets are retained throughout.

The memory oscillator is part of the evolving source neighborhood. Radiation that leaves it into $R$ is accessible; radiation that leaves into $E$ is terminally lost. An excitation can cycle between the source and memory before leaving, so the source alone has a time-nonlocal evolution even though the enlarged source--memory system is Markovian with respect to its external continua. The resulting temporal wavepacket is kept coherently. Treating each emission time as a classical label and averaging those labels would add a dephasing operation absent from this Hamiltonian.

For one rail, write $\kappa=\kappa_R+\kappa_E$, $\eta=\kappa_R/\kappa$. The source and memory amplitudes in the rotating frame satisfy
\begin{equation}
 \dot c=-\gamma c/2-iga,\qquad \dot a=-\kappa a/2-igc,
 \qquad c(0)=1,\quad a(0)=0.
 \label{eq:amplitudes}
\end{equation}
The full collected wavepacket is $\sqrt{\kappa_R}a(s)$, whose integrated norm gives $p_R(t)$. These amplitudes arise from the vacuum quantum model, rather than a classical fluctuating force. Their exact asymptotic yield is
\begin{equation}
 p_R(\infty)=\frac{\eta\Gamma}{\Gamma+\gamma}\frac{\kappa}{\kappa+\gamma},
 \qquad \Gamma=4g^2/\kappa .
 \label{eq:yield}
\end{equation}
Holding the on-resonance rate $\Gamma$, source loss $\gamma$ and split $\eta$ fixed, longer memory changes the emission timescale and lowers collection when the memory lies before the collected port. Filtering the loss port instead changes the branching differently (Extended Data Fig.~\ref{fig:ed1}b). These effects motivate comparisons at equal output resources.

\begin{figure}[!htbp]
\centering\includegraphics[width=\textwidth]{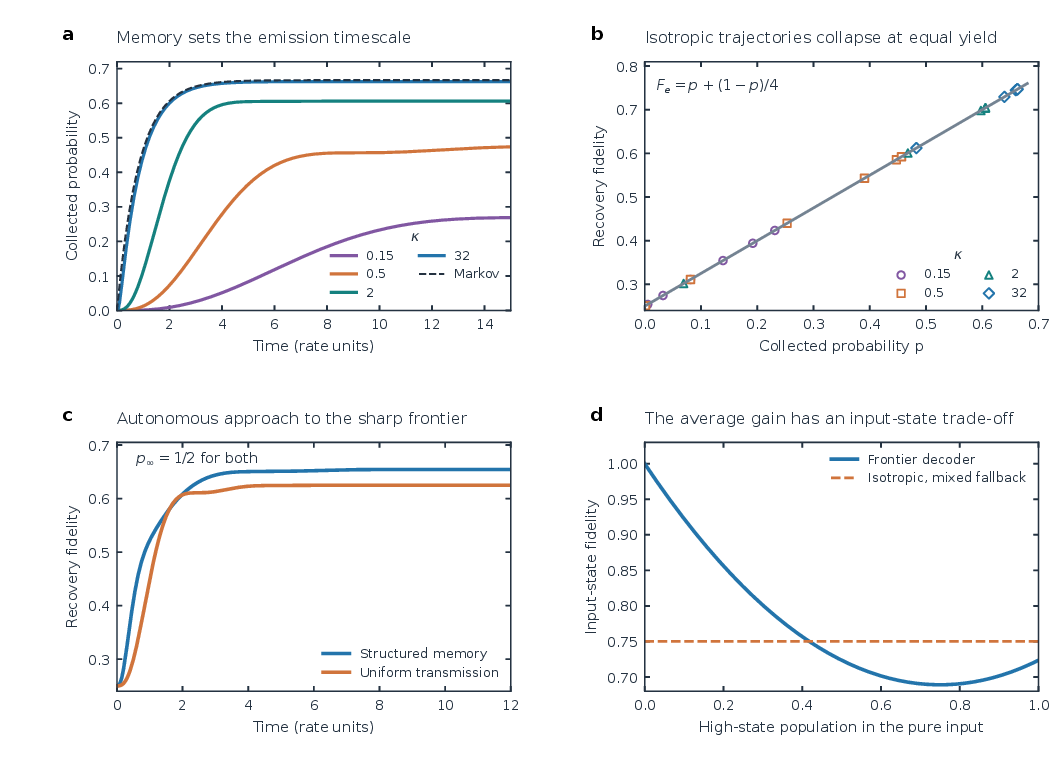}
\caption{\textbf{Autonomous dynamics resolves the role of memory.} \textbf{a}, Collected probability at $\Gamma=1$, $\gamma=0.2$, $\eta=0.8$ and the indicated bandwidths $\kappa$. The dashed curve is the matched-rate Markov limit. \textbf{b}, Identical-rail trajectories collapse onto one fidelity curve when compared at equal collected probability. Symbols sample different times and bandwidths; the line is the exact isotropic expression. \textbf{c}, Unequal fixed bandwidths approach the sharp qubit frontier, while the equal-transmission comparator approaches $0.625$. Both collect one-half asymptotically; their time profiles differ. Parameters appear in Extended Data Table~\ref{tab:params}. \textbf{d}, The frontier decoder's fidelity for individual pure inputs, compared with the isotropic channel using a mixed vacuum fallback. The higher average permits a trade-off among input states. Time is expressed in the chosen inverse-rate unit.}\label{fig:dynamics}
\end{figure}

With identical rails, every trajectory obeys $F_e=p_R+(1-p_R)/k^2$, independently of bandwidth (Fig.~\ref{fig:dynamics}a,b). At matched collected probability, the apparent memory advantage disappears. With zero direct source loss, equal total emission and equal collection likewise produce the same channel up to a receiver unitary. These controls identify what an improvement must change: the transmission spectrum across encoded states.

Figure~\ref{fig:dynamics} separates timing from transmission structure. Panel~a shows that bandwidth changes the collection history; panel~b removes that timing difference by comparing fidelity at equal collected probability. The identical-rail trajectories coincide. Unequal rails in panels~c,d leave a recovery difference after asymptotic yield is matched. The fixed couplings create a state-dependent transmission spectrum; the mechanism does not require a receiver-controlled reversal.

Unequal static bandwidths realize precisely that change. At $\Gamma=4$, $\gamma=1$ and $\eta=1$, choosing $(\kappa_h,\kappa_l)=(9.472136,0.527864)$ produces Eq.~\eqref{eq:qubit}; the corresponding couplings are $(g_h,g_l)=(3.077684,0.726543)$. The equal-transmission comparator has $\kappa=5/3$ and $g=\sqrt{5/3}$ on both rails (Extended Data Table~\ref{tab:params}). Both sources fully decay and collect half of the mean initial carrier energy, while approaching different fidelities autonomously (Fig.~\ref{fig:dynamics}c). Equal carrier energy uses degenerate narrowband modes; the wavepacket spectra and time profiles remain distinct. The advantage is thus an attainable effect of structured transmission, rather than a universal benefit of increasing memory time.

Reference mutual information tracks correlations in the radiation, inaccessible output and remaining source plus memory (Extended Data Fig.~\ref{fig:ed1}a). This accounting retains coherent superpositions of emission times and makes the access restriction explicit. It also shows why source relaxation alone is insufficient to infer successful recovery.

\FloatBarrier
\section*{A radiation-only coding benchmark}
A coding test asks a different question from average single-use fidelity. For independently prepared emitters, consider $\rho_x=\operatorname{diag}(x,1-x)$ and actual collected probability $p_x=x\lambda_h+(1-x)\lambda_l$. The coherent information is the collected-output entropy minus the complementary-output entropy. Its positivity supplies an achievable unassisted rate\cite{Filippov,Devetak}. For the frontier spectrum, $x=0.1558153$ gives
\begin{equation}
 I_c=0.08607388\ \text{qubits per independent emission},\qquad
 p_x=0.34607593 .
 \label{eq:rate}
\end{equation}
An isotropic channel matched to this actual collection has zero unassisted capacity, because its transmission is below one-half (Fig.~\ref{fig:frontier}c,d). Matching $p_x$, rather than the maximally mixed input's yield, is essential for this comparison. Both preparations still contain one initial excitation and decay completely.

The capacity effect of state-dependent erasure has established precedents\cite{Filippov}; here it tests the autonomous implementation under a controlled energy and access comparison. The quoted rate is a lower bound for coding across independent preparations. It complements the exact single-use frontier and makes clear why an emitted-energy budget alone does not determine useful quantum information.

The coding comparison uses the yield of its chosen input ensemble. Bias towards the weakly transmitted state makes that yield smaller than the maximally mixed value. Matching this actual yield gives the isotropic comparator the same mean collected carrier energy (Fig.~\ref{fig:frontier}d). Independent preparation of the source and reservoir defines each channel use; a repeated unreset memory would require a different coding analysis.

\section*{Observable recovery at equal yield}
The qubit comparison can be expressed through three reference--output correlations after the specified deterministic decoder. With Pauli operators $X,Y,Z$ on the reference and decoded output,
\begin{equation}
 F_e=\frac{1+\langle X_AX_C\rangle-\langle Y_AY_C\rangle+
 \langle Z_AZ_C\rangle}{4}.
 \label{eq:observable}
\end{equation}
Failed collection events are assigned the decoder's vacuum fallback and retained. For transmission eigenvalues $\lambda_h\ge\lambda_l$, the three correlations are $\sqrt{\lambda_h\lambda_l}$, $-\sqrt{\lambda_h\lambda_l}$ and $\lambda_h$. The optimal structured channel therefore has reduced transverse coherence but an increased longitudinal correlation compared with equal transmission; their combination gives the gain in Eq.~\eqref{eq:qubit}. Both channels have mean collection one-half. This identifies the measurable change underlying the optimized fidelity, including the information carried by failure. Supplementary Information S16 derives the correlations from the decoded channel and gives independent matrix checks. Such a test presumes a prepared reference and calibrated radiation-mode processing in a controlled implementation. An overlap above the separable-state threshold certifies surviving entanglement, but does not by itself certify positive unassisted quantum capacity.

\section*{Discussion}
The central physical question is whether an environmental mechanism changes the channel available to an outside observer. The fixed-yield frontier makes that question quantitative without granting access to the source or terminal environment. It identifies the best deterministic recovery over the stated one-excitation channel class, and a fixed-coupling emitter approaches the qubit optimum. The comparison includes failed collection events, so its advantage cannot be obtained by discarding unsuccessful trials. The optimized code theorem separately identifies the best subspace of a given source transmission map.

The origin of the gain is important for interpreting the result. Unequal transmissions reduce the coherence in successful emission but make the vacuum event more informative about the input basis state. The exact frontier balances these effects. It improves average entanglement fidelity while permitting a trade-off among individual inputs, as Fig.~\ref{fig:dynamics}d shows. Its large-code advantage decreases as $p/k^3$. This scaling confines the claimed benefit and prevents a small-code result from being interpreted as a fixed recovery fraction for a macroscopic black-hole state.

The matched-yield controls identify a frequent ambiguity in memory-assisted recovery. A longer-lived intermediate mode changes when an excitation escapes and how often it is lost. Comparing at equal time can therefore conflate a change in emission history with a change in information transmission. Identical rails collapse onto the same fidelity curve after collection is matched. The unequal-rail construction retains a difference at the same asymptotic yield because it changes the transmission spectrum across the code. The experimentally accessible distinction is consequently in the complete reference--output correlations, rather than in source coherence or photon counts alone.

The spectral criterion supplies a complementary test that does not require solving the source's interacting dynamics. Within its bath assumptions, a collected output dominated in every spectral direction by terminal loss is antidegradable. No choice of internal interactions or memory inside that factorization can create unassisted quantum capacity. Violating the ordering removes this sufficient obstruction but does not establish recovery. This gives microscopic model building a concrete target: identify a physically justified change in the accessible output map, then test it. A proposed mechanism that changes only a common memory envelope while preserving the ordered output partition remains subject to the same obstruction.

For evaporation, this distinguishes global information bookkeeping from delivery to an observer. The Sachdev--Ye--Kitaev calculation provides a source with an established low-energy gravitational connection\cite{Maldacena,Kitaev,MaldacenaYang}; the matrix criterion applies to its specified coupling independently of the finite-source spectral diagnostic. The sharp frontier and its attainment belong to the separate one-excitation realization. Combining these statements does not identify the emitter with four-dimensional black-hole microphysics. Their role is to constrain the interpretation of an openness-based mechanism once its outgoing-field map and access conditions have been physically justified, including backreaction and the distinction between returning fields and terminally inaccessible radiation.

The observable correlation test makes the operational comparison reproducible in a controlled implementation, in the spirit of verified scrambling measurements\cite{Landsman}. The positive-rate example adds a different consequence: at matched actual output yield, structured transmission can support unassisted coding where uniform transmission cannot. That example uses independent preparations and a biased ensemble; its rate statement cannot be inferred from a single Bell overlap. The combination of a sharp finite-use benchmark, a microscopic exclusion criterion and explicit observables is useful beyond evaporation, wherever radiation or a quantum-network output is only partly collected. It separates improvements due to the accessible channel from improvements purchased through extra access or a larger output budget.

\section*{Conclusion}
Quantum memory assists an outside observer only through its effect on the accessible radiation channel. We have established an attainable recovery frontier at fixed mean yield, determined the optimal permitted subspace encoding, and constructed autonomous dynamics approaching the qubit limit. A broader microscopic spectral condition rules out unassisted transmission when terminal loss dominates every collected direction. Joint reference--output correlations distinguish the recovery gain without increasing mean collection, while matched-yield controls identify when memory changes only the emission history. These results provide explicit tests for proposed quantum-openness mechanisms in evaporation and a constrained route for identifying which changes could make information available in escaping radiation.


\clearpage
\section*{Methods}
\textbf{Quantum noise and spectral reduction.}
The continuum convention is $[b_{\alpha r}(\omega),b_{\beta s}^\dagger(\omega')]=\delta_{\alpha\beta}\delta_{rs}\delta(\omega-\omega')$. Direct integration of Eq.~\eqref{eq:microscopic} gives
\begin{equation}
 \bm b_\alpha(\omega,t)=e^{-i\omega t}\bm b_\alpha(\omega,0)
 -i\int_0^t ds\,e^{-i\omega(t-s)}G_\alpha(\omega)\bm O(s).
 \label{eq:bath}
\end{equation}
For one Hermitian source operator and $B=\int d\omega\sqrt{J(\omega)}(b_\omega+b_\omega^\dagger)$, the input-field commutator and symmetric correlation are
\begin{align}
 D^{\rm ret}(\tau)&=-i\theta(\tau)\langle[B_0(\tau),B_0(0)]\rangle
 =-2\theta(\tau)\int_0^\infty d\omega J(\omega)\sin\omega\tau,\nonumber\\
 N(\tau)&=\tfrac12\langle\{B_0(\tau),B_0(0)\}\rangle
 =\int_0^\infty d\omega J(\omega)\coth(\beta\omega/2)\cos\omega\tau .
 \label{eq:noise}
\end{align}
The vacuum limit is $\beta\to\infty$. Eliminating the bath retains the operator commutator and the same spectral density in response and noise. The full bath solution retains emitted phases that a source-only reduced state would discard.

Write $G=(G_R,G_E)^T$. Its polar decomposition supplies coupled bath modes; orthogonal modes stay in their input state. Diagonalizing the collected compression of this isometry yields independent attenuators of transmissions $\eta_j(\omega)$ after a passive receiver rotation. An arbitrary non-Gaussian coupled output is allowed. When every $\eta\le1/2$, attenuation of the inaccessible output by $\eta/(1-\eta)$ simulates the receiver. Composition with arbitrary source dynamics preserves antidegradability. The norm of the sum of two maximally entangled target projectors is $1+1/k$, proving the fidelity ceiling. The frequency-dependent proof concerns the complete outgoing channel; discarding late radiation also preserves its obstruction.

Identical thermal occupations in all degenerate input modes preserve the bath rotation and the same antidegradability result. The thermal statement requires this shared covariance; unequal temperatures, squeezing and initial source--bath correlations change the channel assumptions. For vacuum inputs, define $\delta=\sup_{j,\omega}[\pi/4-\arccos\sqrt{\eta_j(\omega)}]_+$. If the total coupled-output mean photon number is bounded by $\bar n$, clipping the transmissions above one-half gives
\begin{equation}
 \epsilon\le\min\{1,\sqrt{2\bar n(1-\cos\delta)}\},\qquad
 F_\star\le\min\{1,(k+1)/(2k)+\epsilon\}.
 \label{eq:stability}
\end{equation}
Here $\epsilon$ bounds trace distance, including an arbitrary reference. The estimate follows from the overlap of the two beam-splitter dilations (Supplementary Information, S5). It is conservative. An energy version requires a lower frequency cutoff, so that $\bar n\le\bar E/\omega_{\min}$.

\textbf{Autonomous emitter and physical regime.}
The retained source--memory Hamiltonian is
\begin{equation}
 H_{SM}=\omega_0\left(\sum_j|s_j\rangle\langle s_j|+\sum_j a_j^\dagger a_j\right)
 +\sum_jg_j(a_j^\dagger|g\rangle\langle s_j|+\mathrm{h.c.}).
 \label{eq:autonomous}
\end{equation}
Flat external vacuum continua generate memory damping and direct source loss. For one rail,
\begin{equation}
 \dot a=-(i\omega_0+\kappa/2)a-igS^-
 -\sqrt{\kappa_R}b_{R,\rm in}-\sqrt{\kappa_E}b_{E,\rm in},
 \quad b_{R,\rm out}=b_{R,\rm in}+\sqrt{\kappa_R}a .
 \label{eq:langevin}
\end{equation}
The inputs obey $[b_{\alpha,\rm in}(t),b_{\beta,\rm in}^\dagger(t')]=\delta_{\alpha\beta}\delta(t-t')$ and vacuum correlations. Direct source loss has its own independent quantum input. In the one-excitation sector,
\begin{equation}
 p_R(t)=\int_0^t\kappa_R|a(s)|^2ds,\qquad
 p_E(t)=\int_0^t[\gamma|c(s)|^2+\kappa_E|a(s)|^2]ds,
 \quad |c|^2+|a|^2+p_R+p_E=1 .
 \label{eq:outputs}
\end{equation}
Orthogonal rail labels make $A_T=\operatorname{diag}(p_{Rj})$. The receiver retains all temporal modes and can coherently undo their isometry. This benchmark describes one quantum in a superposition of emission times. The arbitrary-source spectral theorem has no one-excitation restriction.

Rates are expressed in a common inverse-time unit, with $\hbar=k_B=1$ and carrier frequency $\omega_0=1000$. The rotating-wave, narrowband model requires $g,\kappa,\gamma\ll\omega_0$. Main trajectories use $\Gamma=1$, $\gamma=0.2$, $\eta=0.8$ and $\kappa=0.15,0.5,2,32$; $g=\sqrt{\Gamma\kappa}/2$. The main maximum $\kappa/\omega_0$ is $0.032$. The extended bandwidth table reaches $100$ as a formal envelope-model scan. Frontier parameters are in Extended Data Table~\ref{tab:params}; the machine-readable tables retain full precision. Matching yield fixes mean collected carrier energy, rather than every frequency-resolved energy moment or time profile. The autonomous total Hamiltonian includes interaction energy.

\textbf{Matched resources and optimization domain.}
For the deterministic frontier, $p=\Tr(T^\dagger T)/k$ is the collection probability of the maximally mixed code. The channel class consists of all contractions $T$ for one excitation with an orthogonal vacuum flag. Fixing $p$ does not fix collection separately for every pure input, the temporal waveform, or the full frequency-resolved energy flux. The autonomous examples match the mean carrier-energy yield of degenerate narrowband rails. The decoder acts on all collected temporal modes, and every failed event remains in its trace-preserving output.

The subspace theorem optimizes the encoding of a fixed physical map. The fixed-yield frontier instead optimizes its transmission eigenvalues under a trace constraint. These are separate optimizations. The coding benchmark uses the actual yield of its nonuniform input, and optimizes a single-letter lower bound rather than the regularized quantum capacity. This separation prevents a gain in one task from being assigned to another.

\textbf{Recovery, coding and information accounting.}
The exact decoder undoes the polar isometry on collection and prepares a minimum-eigenvalue state on vacuum. For a pure qubit input with high-state population $x$, its fidelity is
\begin{equation}
 f(x)=[\sqrt{\lambda_h}x+\sqrt{\lambda_l}(1-x)]^2+
 [1-\lambda_hx-\lambda_l(1-x)](1-x).
 \label{eq:inputf}
\end{equation}
The isotropic comparator in Fig.~\ref{fig:dynamics}d uses a maximally mixed vacuum fallback and has input-independent fidelity $0.75$. The average pure-state fidelity is $(kF_e+1)/(k+1)$, relating the frontier to the familiar distinction between disturbance and information gain\cite{Banaszek}. The code optimum covers isometric subspaces of the allowed source sector.

For the diagonal input used in the coding test, direct output and complementary density matrices give
\begin{align}
 I_c(x)&=H(x\lambda_h,(1-x)\lambda_l,1-p_x)\nonumber\\
 &\quad-H(x(1-\lambda_h),(1-x)(1-\lambda_l),p_x),
 \label{eq:coherent}
\end{align}
where $H$ is Shannon entropy in bits. The coding theorem yields $Q\ge\max\{0,I_c(x)\}$ for independent channel uses\cite{Devetak}. The isotropic comparator at actual yield $p_x$ is realized with $\Gamma=4$, $\gamma=1$, $\eta=1$, $\kappa=p_x/(0.8-p_x)$ and $g=\sqrt\kappa$. This is an achievable asymptotic rate, rather than a finite-block decoding construction. For any output group $X$ with orthogonal one-particle probabilities $p_j$, the initial maximally entangled code gives
\begin{equation}
 I(A:X)=\log_2k+H(\{p_j/k\},1-\bar p)
 -H(\{(1-p_j)/k\},\bar p),\quad \bar p=k^{-1}\sum_jp_j .
 \label{eq:mutual}
\end{equation}
Purity of the global state establishes this expression for collected radiation, inaccessible radiation and retained source plus memory separately.

\textbf{Microscopic spectra and gravitational regime.}
Independent Gaussian SYK couplings have the variance in Eq.~\eqref{eq:syk}. Jordan--Wigner Majoranas implement the canonical algebra, with $O=2i\chi_1\chi_2$, $J=1$, $\beta=4$, and seeds $20261023+100N+r$ for $r=0,1,2$. The thermal trace includes both parity sectors; physical codes are restricted to a permitted sector. The positive-frequency probe spectrum is
\begin{equation}
 J_{\rm bath}(\omega)=0.02\left(\frac{\omega}{1+\omega}\right)^2
 \frac{0.3^2}{(\omega-0.35)^2+0.3^2},\qquad \omega>0 .
 \label{eq:regulator}
\end{equation}
For thermal probabilities $p_n$, exact source eigenstates give the leading initial emission power
\begin{equation}
 P_\alpha=2\pi\sum_{E_n>E_m}(E_n-E_m)J_\alpha(E_n-E_m)
 p_n|\langle m|O|n\rangle|^2 .
 \label{eq:power}
\end{equation}
These finite-size calculations retain all microscopic transition weights. They test the specified thermal power observable; the phase-sensitive recovery results follow from the channel derivations. Figure~\ref{fig:spectral}b bins the line weights in intervals $0.025J$; the unbinned data are retained in the reproducibility files. Figure~\ref{fig:spectral}c shows three realizations per size, sufficient to test the algebraic partition identity rather than estimate a thermodynamic limit.

In the large-$N$ conformal window, the fixed-index nonsinglet bilinear has leading dimension $1/2$ from factorization of two fermion correlators. The super-Ohmic probe has dimension $3/2$, making the weak boundary coupling dimension $-1$. The source soft action is $I_{\rm soft}=-C\int du\{\tan[\pi f(u)/\beta],u\}$ with $C\propto N/J$\cite{Maldacena,Kitaev}. This supplies the controlled infrared interpretation of the probe; extending a model to complete gravitational evaporation additionally requires its time-dependent backreaction and bulk operator dictionary. Equation~\eqref{eq:nogo} remains a Hamiltonian-level statement outside the conformal approximation.

\subsection*{Unrestricted decoder programs and numerical certificates}
For Kraus matrices $K_\mu:\mathbb C^k\to\mathbb C^r$, let
\begin{equation}
 v_\mu=\sum_{i=1}^k\sum_{a=1}^r(K_\mu)_{ai}^*|i\rangle|a\rangle,
 \qquad M=k^{-2}\sum_\mu v_\mu v_\mu^\dagger.
 \label{eq:sdpmatrix}
\end{equation}
In this output--input ordering for the recovery Choi matrix $J_D$, the independent primal and dual problems are
\begin{equation}
 \max_{J_D\succeq0,\,\operatorname{Tr}_{\mathbb C^k}J_D=I_r}\operatorname{Tr}(MJ_D),
 \qquad
 \min_{Y=Y^\dagger,\,I_k\otimes Y\succeq M}\operatorname{Tr}Y .
 \label{eq:sdp}
\end{equation}
The Choi matrix is unnormalized, with trace $r$. This convention fixes all transposes and factors of $k$ in the objective. The optimizer acts on the whole output, including failure flags. No heralding or restriction to unitary recovery is imposed. The implementation uses CVXPY and CLARABEL, with absolute gap, relative gap and feasibility tolerances $2\times10^{-8}$ and at most 150 iterations. Raw objectives and solver statuses are saved alongside the repaired certificates.

For a numerical lower certificate, negative eigenvalues of $J_D$ are clipped to zero and its input partial trace $T$ is computed. Congruence by $I_k\otimes T^{-1/2}$ restores trace preservation. The tested instances have nonsingular $T$. For an upper certificate, $Y$ is shifted by $[\max(0,-\lambda_{\min}(I_k\otimes Y-M))+10^{-12}]I_r$. The resulting primal--dual interval checks the analytical value, subject to floating-point accuracy. This is distinct from replacing a failed solver status by an assumed optimum. Exact results in the main text have analytical constructions and upper bounds in the Supplementary Information.

\textbf{Numerical verification.}
Population-generator exponentiation is checked against complex amplitude exponentiation plus Gauss quadrature. Unrestricted primal and dual semidefinite recovery programs\cite{Fletcher} test twelve cases, including complex rectangular maps and dimensions two to four. Raw and feasibility-repaired objectives are retained. Two complex dual instances trigger an explicit SCS fallback after a CLARABEL accuracy warning; final statuses are optimal. The maximum discrepancy from the analytical fidelity is $1.89\times10^{-8}$, and the dynamical cross-check differs by at most $8.66\times10^{-15}$. These are floating-point tests; general claims have separate analytical proofs.

For the dynamical cross-check, define $u=|c|^2$, $v=|a|^2$, $z=\operatorname{Im}(c^*a)$ and $\bm x=(u,v,z,p_R,p_E)^T$. The real generator used in the production tables is
\begin{equation}
 \dot{\bm x}=\begin{pmatrix}
 -\gamma&0&2g&0&0\\0&-\kappa&-2g&0&0\\
 -g&g&-(\gamma+\kappa)/2&0&0\\
 0&\eta\kappa&0&0&0\\\gamma&(1-\eta)\kappa&0&0&0
 \end{pmatrix}\bm x,
 \qquad \bm x(0)=(1,0,0,0,0)^T .
 \label{eq:generator}
\end{equation}
Its conserved row combination is $u+v+p_R+p_E=1$. The independently implemented check exponentiates the complex $2\times2$ amplitude matrix and integrates the outgoing intensities with 100-point Gauss--Legendre quadrature. These formulations share physical parameters but not the numerical propagation or integration algorithm. The closed-form infinite-time yield is checked separately, so an apparently converged finite-time trace is not used as its definition.

The coherent-information calculation forms both output and complementary density matrices from the Kraus operators, diagonalizes them, and compares their entropy difference with Eq.~\eqref{eq:coherent}. A bounded scalar optimization is checked against a uniform grid of 2,001 input populations for each tested spectrum. Entropies are in bits, with $0\log_2 0=0$. Random complex isometries test the subspace result but do not establish its global optimum; eigenvalue interlacing supplies that proof. Similarly, the fixed-yield frontier is compared with an independent one-variable optimization after equalizing the larger transmissions.

Every plotted curve is generated from a named CSV table. Numerical uncertainty in deterministic results is assessed through residuals, conservation and independent algorithms, rather than artificial statistical error bars. The three SYK realizations per size test an exact proportional-spectrum identity and are not used for a claim about disorder-averaged scaling. The reproducibility files record the Python and dependency versions, fixed seeds and full-precision parameters. Supplementary Fig.~S1 compares independent decoder and dynamical calculations and displays their residuals.

Further checks include 606 independent scalar frontier optimizations, 2,000 random contraction spectra, canonical anticommutators, thermal detailed balance, conservation, positivity and reference-entangled channel tests. Coherent information is checked against full Kraus output and complementary density matrices, with optimization compared against 2,001 populations for each of 147 transmission spectra. All 28 original verification groups pass. An additional 110 qubit cases compare the explicit decoded density matrix, Pauli correlations and exact Bell overlap. A fresh-directory rerun reproduces the original 16 CSV tables with no numerical differences in the pinned-environment rerun. This tests reproducibility with the supplied algorithms, complementing the independent numerical formulations. Code, seeds, tolerances and all residuals are retained in the reproducibility files.

	\textbf{Acknowledgment.}
K.A.Y. acknowledges support from the Research Fund for International Young
Scientists of the NSFC under Grant No.~KYZ04Y22050, Zhejiang Normal University
research funding under Grant No.~ZC304021914, and the Zhejiang Province
postdoctoral research project under Grant No.~ZC304021952.

\textbf{Author contributions.}
K.A.Y. conceived the study, developed the model, performed the analytical and
numerical calculations, analyzed the results, and wrote the manuscript.

\textbf{Data availability.}
The numerical data supporting this study are available from the corresponding author upon reasonable request. No observational or experimental data are used.

\textbf{Code availability.}
The source code used for the calculations and figures is available from the corresponding author upon reasonable request.


\begingroup\renewcommand{\bibfont}{\fontsize{8.8}{10.3}\selectfont}

\endgroup
\clearpage
\renewcommand{\thefigure}{\arabic{figure}}
\setcounter{figure}{0}
\renewcommand{\theHfigure}{ED\arabic{figure}}
\section*{Extended Data}
\renewcommand{\thetable}{\arabic{table}}
\makeatletter\renewcommand{\fnum@figure}{Extended Data Fig.~\thefigure}\renewcommand{\fnum@table}{Extended Data Table~\thetable}\makeatother
\begin{figure}[h!]
\centering\includegraphics[width=\textwidth]{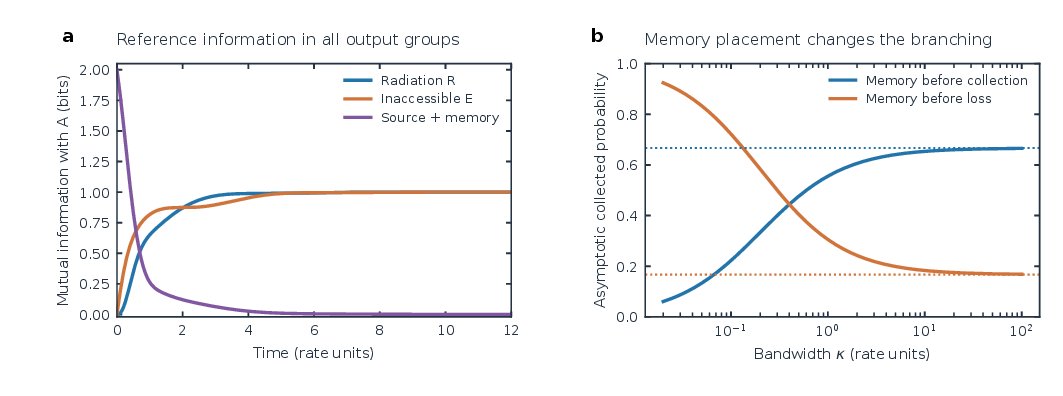}
\caption{\textbf{Information accounting and the location of quantum memory.} \textbf{a}, Mutual information of the reference $A$ with collected radiation $R$, terminal inaccessible output $E$, and the remaining source plus memory for the structured qubit realization. The code initially contains one excitation and is maximally entangled with $A$; all external inputs are vacuum. Full temporal wavepackets are retained within each output group. Correlations are calculated from the global quantum isometry, rather than inferred from the source population. At finite time the three plotted mutual informations need not add to a conserved total, because correlations can be shared among three output groups. \textbf{b}, Asymptotic collection when the memory filters the collected port or the inaccessible port. Both curves use $\Gamma=1$ and $\gamma=0.2$. Before collection, $\eta=0.8$ of memory leakage reaches $R$ and direct decay enters $E$; before loss, all memory leakage enters $E$ and direct decay reaches $R$. Dotted lines show the respective matched-rate Markov limits. This comparison changes the port assignment and hence the collection conditions: it diagnoses placement, rather than demonstrating a recovery gain at matched yield. Supplementary Information S9 gives the channel definitions; the source tables retain the full bandwidth scan.}\label{fig:ed1}
\end{figure}
\begin{table}[!htbp]
\caption{\textbf{Autonomous parameters of the matched-yield comparison.} Both rails have $\Gamma=4$, $\gamma=1$, $\eta=1$ and carrier frequency $\omega_0=1000$. Rates are in the common inverse-time unit. Static parameters fix the asymptotic transmission eigenvalues; the memory correlation time is $2/\kappa$. Rounded values are shown; full precision is retained in \texttt{matched\_yield\_parameters.csv}. The additional isotropic comparator for the coding ensemble is given in Methods.}
\begin{ruledtabular}\begin{tabular}{lrrrr}
Case and rail & $p_R(\infty)$ & $\kappa$ & $g$ & $2/\kappa$\\
Isotropic, either & $0.500000$ & $1.666667$ & $1.290994$ & $1.200000$\\
Frontier, high & $0.723607$ & $9.472136$ & $3.077684$ & $0.211146$\\
Frontier, low & $0.276393$ & $0.527864$ & $0.726543$ & $3.788854$\\
\end{tabular}\end{ruledtabular}\label{tab:params}
\end{table}
\end{document}


\title{Supplementary Information: Quantum Recovery from Black-Hole Radiation Beyond the Emission Yield}
\author{Kashif Ammar Yasir}
\affiliation{Department of Physics, Zhejiang Normal University, Jinhua 321004, China}
\affiliation{Zhejiang Institute of Photoelectronics, Jinhua 321004, China}
\date{28 October 2026}
\maketitle

\section*{S1. Definitions, assumptions and the role of each model}
The reference $A$ is initially maximally entangled with a code $C$ of dimension $k$. A source encoding is an isometry $W:C\to S$ inside a permitted superselection sector. The observer receives only $R$. The source remainder, memory not yet emitted, and terminal inaccessible outputs are traced out. All decoders are deterministic CPTP maps, with no conditioning on emission, bath readout, initial shared entanglement, source control or echo. The entanglement fidelity is the squared overlap with $\Phi_k$; trace distance is $D(\rho,\sigma)=\|\rho-\sigma\|_1/2$.

There are two distinct models. First, the spectral theorem assumes an arbitrary source with a linear coupling to Gaussian bosonic bath fields, a source-independent initial bath, and well-defined outgoing modes. Its vacuum result allows arbitrary source interactions, multi-quantum radiation and memory spectra. The thermal extension requires the same thermal occupation in all degenerate incoming modes. Second, the sharp recovery frontier is for the generalized erasure channel produced by a one-excitation source. Its autonomous realization uses a rotating-wave narrowband envelope model and vacuum inputs. This realization resolves the one-excitation output channel. Finite-temperature multi-excitation emission requires a larger channel space; the arbitrary-source spectral theorem already admits general emission multiplicity.

The finite SYK calculations are exact source spectra and matrix elements, used for a leading weak-coupling thermal emission diagnostic. They test a finite-source initial emission observable. The spectral obstruction holds independently of these numerical diagnostics. The gravitational interpretation of SYK is its established infrared connection to nearly AdS$_2$ gravity\cite{Maldacena,Kitaev}; the extension to four-dimensional evaporation requires its microscopic Hamiltonian and outgoing operator dictionary.

\section*{S2. Eliminating a quantum bath without replacing its state by noise}
For a scalar Hermitian source operator $O$, take
\begin{equation}
 H=H_S+\int_0^\infty d\omega\,\omega b_\omega^\dagger b_\omega
 +O\int_0^\infty d\omega\,\sqrt{J(\omega)}(b_\omega+b_\omega^\dagger),
 \quad [b_\omega,b_{\omega'}^\dagger]=\delta(\omega-\omega').
\end{equation}
Form factors are regulated so the Hamiltonian and integrals are defined. Exact integration gives
\begin{align}
 b_\omega(t)&=e^{-i\omega t}b_\omega(0)
 -i\sqrt{J(\omega)}\int_0^t ds\,e^{-i\omega(t-s)}O(s),\\
 B(t)&=B_0(t)+\int_0^t ds\,D^{\rm ret}(t-s)O(s),\\
 B_0(t)&=\int_0^\infty d\omega\,\sqrt{J(\omega)}
 [e^{-i\omega t}b_\omega(0)+e^{i\omega t}b_\omega^\dagger(0)],\\
 D^{\rm ret}(\tau)&=-2\theta(\tau)\int_0^\infty d\omega\,J(\omega)\sin\omega\tau.
\end{align}
For an arbitrary source observable $X$, its Heisenberg equation can therefore be written, with the displayed operator ordering, as
\begin{equation}
 \dot X(t)=i[H_S(t),X(t)]+i[O(t),X(t)]
 \left[B_0(t)+\int_0^t ds\,D^{\rm ret}(t-s)O(s)\right].
\end{equation}
This is generally nonlinear and operator-valued; a linear Langevin equation for the complete SYK source has not been assumed. Initial thermal bath correlations are
\begin{align}
 \langle b_\omega^\dagger b_{\omega'}\rangle&=n_\beta(\omega)\delta(\omega-\omega'),\quad
 \langle b_\omega b_{\omega'}^\dagger\rangle=[n_\beta(\omega)+1]\delta(\omega-\omega'),\\
 \tfrac12\langle\{B_0(t),B_0(s)\}\rangle
 &=\int_0^\infty d\omega\,J(\omega)[2n_\beta(\omega)+1]\cos\omega(t-s),\\
 [B_0(t),B_0(s)]&=-2i\int_0^\infty d\omega\,J(\omega)\sin\omega(t-s).
\end{align}
The commutator and symmetric noise kernel have the same spectral density. Noise strength and damping cannot be varied independently while retaining this microscopic equilibrium bath. The vector formula in the main text follows by retaining the coupling matrices. Crucially, the integrated bath operator also specifies emitted quantum amplitudes and correlations; tracing the source's reduced master equation alone would not supply the radiation channel needed for recovery.

\section*{S3. Proof of the matrix spectral obstruction}
At each frequency set $G=(G_R^T,G_E^T)^T$, $J=G^\dagger G$ and let the inverse square root act only on the support of $J$. Then
\begin{equation}
 C=GJ^{-1/2},\quad C^\dagger C=I,\qquad
 \bm c=C^\dagger\begin{pmatrix}\bm b_R\\\bm b_E\end{pmatrix}.
\end{equation}
Complete $C$ to a unitary matrix. The remaining combinations $\bm d$ are dark: the interaction is $\bm c^\dagger J^{1/2}\bm O+\mathrm{h.c.}$ and contains no $\bm d$. The free bath Hamiltonian is unchanged because the rotation mixes only modes with equal frequency. The initial vacuum is a product vacuum in bright and dark modes, so dark modes remain independent of all source--bright dynamics.

Write $C_R=G_RJ^{-1/2}$, $C_E=G_EJ^{-1/2}$. Since $C_R^\dagger C_R=A$ and $C_E^\dagger C_E=I-A$, a singular-value decomposition of $C_R$ diagonalizes the collection map, with complementary singular values for $E$. Up to passive input and separate output unitaries, the outgoing fields have the beam-splitter form
\begin{equation}
 r_j=\sqrt{\eta_j}\,c_j+\sqrt{1-\eta_j}\,d_j,\qquad
 e_j=\sqrt{1-\eta_j}\,c_j-\sqrt{\eta_j}\,d_j,
 \quad \eta_j\in\operatorname{spec}(A).
\end{equation}
Extra spectator modes can be discarded. This is a statement about a bath rotation, not a Gaussian approximation to the source. A highly non-Gaussian or entangled bright output is allowed.

For a vacuum attenuator $\mathcal L_\eta$, composition multiplies transmissivities:
\begin{equation}
 \mathcal L_x\circ\mathcal L_y=\mathcal L_{xy}.
\end{equation}
For $\eta\le1/2$, choose $x=\eta/(1-\eta)\le1$. The inaccessible output, after an irrelevant phase rotation, is $\mathcal L_{1-\eta}$ of the bright state. Applying $\mathcal L_x$ reproduces the receiver. Applying this construction in each frequency and mode gives an antidegrading channel from $E$ to $R$. The full complementary output also includes the source and any bright-channel environment, so a map that needs only $E$ is sufficient. Precomposition by any source dynamics or encoding preserves this construction.

On the support of $J$,
\begin{equation}
 A\preceq I/2\quad\Longleftrightarrow\quad
 J^{-1/2}(J_E-J_R)J^{-1/2}\succeq0
 \quad\Longleftrightarrow\quad J_R\preceq J_E.
\end{equation}
The null space of $J$ is also a null space of both positive semidefinite $J_R$ and $J_E$, so it creates no exception. This proves the sufficient condition in the main text. It is the microscopic reduction to the established antidegradability of lossy bosonic channels\cite{Caruso,MultiCaruso}; the one-half threshold itself is not new.

The result is formulated for asymptotic outgoing fields. Frequency-dependent rotations need not be causal operations on a short time window. This does not affect the global channel obstruction, or its persistence under discarding late radiation or additional detector loss. It would be inappropriate to interpret the proof as a real-time experimental antidegrader with no delay.

\section*{S4. Fidelity ceiling and identical-temperature extension}
Antidegradability provides a joint output $R_1R_2$ with equal marginals, each equal to the receiver's channel. Apply the same decoder to both. With $P_1=\Phi_{AC_1}\otimes I_{C_2}$ and $P_2=\Phi_{AC_2}\otimes I_{C_1}$,
\begin{equation}
 2F_e=\Tr[(P_1+P_2)\rho_{AC_1C_2}]
 \le\|P_1+P_2\|_\infty=1+1/k.
\end{equation}
To verify the norm, define $|u_j\rangle=|\Phi_k\rangle_{AC_1}|j\rangle_{C_2}$ and $|v_j\rangle=|\Phi_k\rangle_{AC_2}|j\rangle_{C_1}$. They are orthonormal within each family and $\langle u_i|v_j\rangle=\delta_{ij}/k$. The nontrivial eigenvalues of the projector sum are $1\pm1/k$. Thus $F_e\le(k+1)/(2k)$ for every code and decoder. Tensor products of antidegradable channels remain antidegradable, excluding reliable unassisted quantum transmission and giving $Q=0$. These claims exclude pre-shared entanglement, environment assistance and postselection.

If every incoming bath mode of frequency $\omega$ has the same occupation $n(\omega)$, the initial Gaussian covariance is proportional to the identity within that degenerate frequency sector. The bright--dark rotation therefore leaves a product of thermal states. The receiver is a thermal attenuator $\mathcal L_{\eta,n}$, and the physical inaccessible output is its weak complementary channel $\mathcal L_{1-\eta,n}$ up to a phase. Thermal attenuators with the same $n$ obey
\begin{equation}
 \mathcal L_{x,n}\circ\mathcal L_{y,n}=\mathcal L_{xy,n},
\end{equation}
because their noise covariance combines as $x(1-y)(2n+1)+(1-x)(2n+1)=(1-xy)(2n+1)$. The map $x=\eta/(1-\eta)$ again simulates the receiver. Purifying the thermal input only enlarges the true complementary output; it does not invalidate the antidegrader that already acts on the physical inaccessible modes. Different temperatures, squeezed inputs, bath correlations, and initially source-correlated inputs are outside this extension.

\section*{S5. Stability under a small spectral violation}
For vacuum inputs let $\eta_j'=\min(\eta_j,1/2)$ and use the same bright-output channel in the original and clipped models. Write the beam-splitter angles as $\theta_j=\arccos\sqrt{\eta_j}$, $\theta_j'=\arccos\sqrt{\eta_j'}$ and $\delta=\sup_j|\theta_j-\theta_j'|$. Frequency is included in the mode label. In a common dilation, for an arbitrary pure bright--reference state and vacuum dark input, the overlap of the two final pure states is
\begin{equation}
 v=\left\langle\prod_j\cos(\theta_j-\theta_j')^{\hat n_j}\right\rangle.
\end{equation}
The operator is positive. For every occupation configuration,
\begin{equation}
 1-\prod_j\cos(\theta_j-\theta_j')^{n_j}
 \le\sum_j n_j[1-\cos\delta].
\end{equation}
This follows from $1-\prod_lx_l\le\sum_l(1-x_l)$ for $0\le x_l\le1$, applied with repeated factors. Hence $v\ge1-\bar n(1-\cos\delta)$ for a state with mean total occupation at most $\bar n$. For the nontrivial range the pure-state trace distance is
\begin{equation}
 D\le\sqrt{1-v^2}\le\sqrt{2\bar n(1-\cos\delta)}.
\end{equation}
Purifying a mixed bright state and tracing unwanted systems cannot increase this distance. Taking the supremum over permitted reference-entangled inputs gives the stated photon-constrained channel estimate, capped by one. A decoder also contracts trace distance, and a projector expectation changes by at most $D$. Adding this to the clipped channel's no-cloning ceiling proves the recovery bound.

The bound is not asserted to be optimal and can become uninformative at large photon budget. It uses total occupation, not merely energy. If all relevant frequencies satisfy $\omega\ge\omega_{\min}>0$, then $\bar n\le\bar E/\omega_{\min}$ provides an energy version. No such replacement is valid for a continuum with arbitrarily soft photons and only a total-energy bound.

\section*{S6. The generalized erasure channel and an exact decoder}
Suppose the one-excitation isometry has collected amplitude $T$ and complementary amplitude $L$ with $T^\dagger T+L^\dagger L=I$. A pure code state evolves as
\begin{equation}
 |\psi\rangle\longmapsto(T|\psi\rangle)_R|0\rangle_{\bar R}
 +|0\rangle_R(L|\psi\rangle)_{\bar R}.
\end{equation}
Collected one-particle and vacuum sectors are orthogonal. Tracing $\bar R$ yields the generalized erasure channel\cite{Filippov}
\begin{equation}
 \cN_T(\rho)=T\rho T^\dagger+\Tr[(I-T^\dagger T)\rho]|0\rangle\langle0|.
\end{equation}
Source and retained memory excitations are included in $\bar R$, not silently dropped. Orthogonal radiation rails are sufficient for the diagonal examples; no measurement of the rail is assumed in the decoder.

Let $T=U\sqrt{A}$ be its polar decomposition. Input and output unitary changes leave the optimization unchanged, so diagonalize $A$ with eigenvalues $\lambda_i$. The success contribution for decoder Kraus operators $D_l$ is
\begin{equation}
 k^{-2}\sum_l\left|\Tr(D_lT)\right|^2.
\end{equation}
On the polar range write $v_i=((D_lU)_{ii})_l$. Trace preservation implies $\|v_i\|\le1$, since the omitted matrix elements in each column contribute nonnegative norms. Therefore
\begin{equation}
 \sum_l\left|\sum_i\sqrt{\lambda_i}(D_lU)_{ii}\right|^2
 =\left\|\sum_i\sqrt{\lambda_i}v_i\right\|^2
 \le\left(\sum_i\sqrt{\lambda_i}\right)^2.
\end{equation}
Applying $U^\dagger$ on the successful subspace attains equality. A CPTP extension to any unused output subspace can be chosen arbitrarily.

On failure the decoder prepares a density matrix $\sigma$. Its contribution is
\begin{equation}
 k^{-2}\Tr[(I-A)\sigma]\le k^{-2}[1-\lambda_{\min}(A)].
\end{equation}
Choose $\sigma$ to be a minimum-eigenvalue eigenstate. The two output sectors are orthogonal, so these independently optimal operations form a single CPTP decoder. This proves
\begin{equation}
 F_\star=\frac{(\Tr\sqrt A)^2+1-\lambda_{\min}(A)}{k^2}.
\end{equation}
If $A$ is singular, the same result follows by continuity or a partial-isometry extension. At $A=0$, it gives $1/k^2$, the entanglement fidelity of an input-independent replacement. At $A=I$, it gives unity. For $A=pI$ it gives $p+(1-p)/k^2$. The vacuum fallback can be any density matrix in this isotropic case; choosing $I/k$ makes its pure-state fidelity input-independent.

\section*{S7. Proof of the all-subspace-encodings optimum}
For a $d$-dimensional full source, a code $W$ replaces $A$ by $A_W=W^\dagger AW$. Order its eigenvalues as $\mu_1\ge\cdots\ge\mu_k$. The min--max principle gives $\mu_i\le\lambda_i^\downarrow(A)$ for each $i$. The function
\begin{equation}
 f(\bm\mu)=k^{-2}\left[\left(\sum_i\sqrt{\mu_i}\right)^2+1-\min_i\mu_i\right]
\end{equation}
is nondecreasing in every coordinate. Away from degeneracies its numerator has derivative $S/\sqrt{\mu_i}$ if that coordinate is not the minimum and $S/\sqrt{\mu_i}-1$ if it is, with $S=\sum_i\sqrt{\mu_i}$. Both are nonnegative. Continuity handles zeros and changes of the minimum. Consequently $f(\bm\mu)\le f(\lambda_1^\downarrow,\ldots,\lambda_k^\downarrow)$. The top $k$ eigenspace attains all these values at once, proving the optimum.

This is a finite-code theorem. A code restricted by energy or symmetry to a particular subspace must first compress $A$ to that allowed subspace. Arbitrary top eigenvectors across forbidden superselection sectors are not physically permitted encodings. The calculation does not claim optimality among mixed, subsystem, assisted, or probabilistically successful encodings.

\section*{S8. Proof of the sharp mean-yield frontier}
Let $r=kp$ and $m=\min_i\lambda_i$. For fixed $r$ and $m$, concavity gives
\begin{equation}
 \sum_{i=1}^{k-1}\sqrt{\lambda_i}\le
 (k-1)\sqrt{\frac{r-m}{k-1}},
\end{equation}
with equality at $k-1$ equal eigenvalues $b=(r-m)/(k-1)$. The allowed range is
\begin{equation}
 \max(0,r-k+1)\le m\le p.
\end{equation}
The upper bound ensures $b\ge m$ and the lower bound ensures $b\le1$. Apart from the overall factor $k^{-2}$, the objective is
\begin{equation}
 h(m)=1+(k-1)(r-m)+2\sqrt{(k-1)m(r-m)}.
\end{equation}
It is concave on the interval. Its stationary point obeys
\begin{equation}
 -(k-1)+\frac{\sqrt{k-1}(r-2m)}{\sqrt{m(r-m)}}=0,
 \qquad m_0=\frac r2\left[1-\sqrt{\frac{k-1}{k+3}}\right].
\end{equation}
For $k\ge2$, $m_0\le p$. Clipping it to the lower allowed endpoint gives exactly the values $m,b$ in the main text. The spectrum $\{b,\ldots,b,m\}$ itself is a valid contraction, so the bound is attainable throughout $0\le p\le1$.

For an explicit piecewise form let
\begin{equation}
 L_k=\frac{k-1+\sqrt{(k-1)(k+3)}}2,
 \qquad p_c=\frac{k-1}{k(1-t_k)}.
\end{equation}
Then
\begin{equation}
 F_{\max}(k,p)=\begin{cases}
 [1+kpL_k]/k^2,&0\le p\le p_c,\\
 [(k-1)^2+1+2(k-1)\sqrt{kp-k+1}]/k^2,&p_c\le p\le1.
 \end{cases}
\end{equation}
At $p=p_c$ the two expressions agree. At $p=0$ and $p=1$ they give $1/k^2$ and $1$. For fixed $p<1$, the first branch applies at sufficiently large $k$. Expanding $L_k/k=1-k^{-2}+k^{-3}+O(k^{-4})$ gives
\begin{equation}
 F_{\max}-\left[p+\frac{1-p}{k^2}\right]=\frac{p}{k^3}+O(k^{-4}).
\end{equation}
The comparison fixes mean collected probability for the maximally mixed code, not collection probability for every pure input. Using $\int d\psi\,(|\psi\rangle\langle\psi|)^{\otimes2}=(I+\mathrm{Swap})/[k(k+1)]$, the Haar-average pure-state fidelity of a trace-preserving decoded channel is $(kF_e+1)/(k+1)$. This explains why the frontier improves that average while allowing lower fidelity for some states. At the qubit example, the average is $0.7696723315$, compared with $0.75$ for the isotropic channel. No worst-case optimality is claimed for the frontier decoder.

\section*{S9. Autonomous quantum-memory amplitudes and outputs}
Take the source--memory Hamiltonian of the main text and couple each $a_j$ to two independent continua with flat envelope couplings $\sqrt{\kappa_{\alpha j}/(2\pi)}$. Couple each $S^-_j$ to a third, inaccessible vacuum continuum with rate $\gamma_j$. The external continua are Markovian in the narrowband envelope approximation. Eliminating them yields
\begin{equation}
 \dot a_j=-(i\omega_0+\kappa_j/2)a_j-ig_jS^-_j
 -\sum_{\alpha=R,E}\sqrt{\kappa_{\alpha j}}\,b_{\alpha j,\rm in},
 \quad \kappa_j=\kappa_{Rj}+\kappa_{Ej}.
\end{equation}
The minus sign of the input is a phase convention compatible with $b_{\rm out}=b_{\rm in}+\sqrt\kappa a$. These operators have vacuum two-point functions and nonzero canonical commutators. The direct loss likewise has a quantum input. They are not classical stochastic drives.

In the one-excitation sector the exact no-output amplitudes of this envelope model satisfy
\begin{equation}
 \frac{d}{dt}\begin{pmatrix}c\\a\end{pmatrix}
 =\begin{pmatrix}-\gamma/2&-ig\\-ig&-\kappa/2\end{pmatrix}
 \begin{pmatrix}c\\a\end{pmatrix},\qquad(c(0),a(0))=(1,0).
\end{equation}
Eliminating $a$ gives
\begin{equation}
 \dot c(t)=-\frac\gamma2c(t)-g^2\int_0^t ds\,e^{-\kappa(t-s)/2}c(s).
\end{equation}
This memory equation for an amplitude is accompanied by the operator noise obtained by convolving the vacuum inputs with the same decaying kernel in the full Heisenberg equation. It is not a replacement of the quantum state by a stochastic trajectory. The memory correlation time is $2/\kappa$; it need not imply information backflow under every proposed definition of non-Markovianity. The pseudomode connection is established in Refs.~\cite{Garraway,Pleasance}.

Define $u=|c|^2$, $v=|a|^2$, $z=\operatorname{Im}(c^*a)$ and integrated output probabilities $p_R,p_E$. They obey
\begin{equation}
 \frac{d}{dt}\begin{pmatrix}u\\v\\z\\p_R\\p_E\end{pmatrix}=
 \begin{pmatrix}
 -\gamma&0&2g&0&0\\
 0&-\kappa&-2g&0&0\\
 -g&g&-(\gamma+\kappa)/2&0&0\\
 0&\eta\kappa&0&0&0\\
 \gamma&(1-\eta)\kappa&0&0&0
 \end{pmatrix}
 \begin{pmatrix}u\\v\\z\\p_R\\p_E\end{pmatrix}.
 \label{eq:population}
\end{equation}
The initial vector is $(1,0,0,0,0)^T$. Differentiating $u+v+p_R+p_E$ gives zero. The emitted probabilities increase monotonically, even if excitation returns from memory to source. This is physically different from applying a deliberate reversal.

For $g,\kappa>0$ and $\gamma\ge0$, the retained amplitudes vanish asymptotically. Integrate the first three equations and set $U=\int u\,dt$, $V=\int v\,dt$, $Z=\int z\,dt$. Then
\begin{equation}
 \gamma U-2gZ=1,\quad \kappa V+2gZ=0,\quad
 g(V-U)-\frac{\gamma+\kappa}{2}Z=0.
\end{equation}
Solving gives
\begin{equation}
 \eta\kappa V=
 \frac{\eta\,4g^2\kappa}{(\gamma+\kappa)(4g^2+\gamma\kappa)}.
\end{equation}
At fixed $\Gamma=4g^2/\kappa$, the $\kappa\to\infty$ limit yields decay rate $\gamma+\Gamma$ and
\begin{equation}
 p_R^{\rm Markov}(t)=\frac{\eta\Gamma}{\Gamma+\gamma}
 [1-e^{-(\Gamma+\gamma)t}].
\end{equation}
The bandwidth comparison holds the on-resonance rate fixed; it does not hold $g$ or the entire spectral density fixed. No claim of identical microscopic resources beyond the quantities stated is made.

For any target $0<p<\eta\Gamma/(\Gamma+\gamma)$ with $\gamma>0$, the asymptotic yield can be engineered by
\begin{equation}
 \kappa=\frac{\gamma p}{\eta\Gamma/(\Gamma+\gamma)-p},\qquad
 g=\sqrt{\Gamma\kappa/4}.
\end{equation}
This formula generates the matched-yield parameter table in Extended Data Table~1. The emitter is autonomous after its initial preparation, but its static couplings are chosen to attain the specified transmission spectrum. It is a realization of a bound, not a claim that a natural black hole chooses that optimum. A finite carrier-to-bandwidth ratio controls the envelope approximation; exact equality of photon yield does not certify equality of the full physical emitted-energy spectrum.

The placement control in Extended Data Fig.~1b uses two explicitly different output partitions. With memory before collection, the collected yield is $\eta\kappa V$ with $\eta=0.8$, while direct decay enters $E$. With memory before loss, all memory leakage enters $E$ and direct source decay enters $R$. Conservation then gives
\begin{equation}
 p_R^{\rm loss\ filter}(\infty)=\gamma U
 =1-\frac{4g^2\kappa}{(\gamma+\kappa)(4g^2+\gamma\kappa)}.
\end{equation}
At $\Gamma=1$ and $\gamma=0.2$, the respective Markov limits are $0.8/1.2$ and $0.2/1.2$. These curves illustrate sensitivity to placement and output labeling; the equal-yield comparisons are the separate frontier examples.

\section*{S10. Keeping the reference and the inaccessible information}
For initially maximally entangled code and reference, the global one-excitation state is
\begin{equation}
 |\Psi(t)\rangle=\frac1{\sqrt k}\sum_j|j\rangle_A\left[
 c_j(t)|s_j,0,0,0\rangle+a_j(t)|g,1_j,0,0\rangle
 +|g,0,f_{Rj}(t),0\rangle+|g,0,0,f_{Ej}(t)\rangle\right].
\end{equation}
The ket entries denote source, memory, collected radiation and inaccessible radiation. Output wavepackets are unnormalized, with squared norms $p_{Rj}$ and $p_{Ej}$. Direct and memory-mediated inaccessible emissions occupy separate orthogonal ports and are included together in $f_{Ej}$. Each rail satisfies $|c_j|^2+|a_j|^2+p_{Rj}+p_{Ej}=1$. Source--memory branches can be combined into an orthogonal retained branch of norm squared $q_j=|c_j|^2+|a_j|^2$.

For any one of these groups $X$, its reduced spectrum is $\{p_j/k\}_{j=1}^k$ together with vacuum weight $1-\bar p$. The complement of $AX$ has spectrum $\{(1-p_j)/k\}_{j=1}^k$ together with weight $\bar p$. Purity of the global state and $S(A)=\log_2 k$ give the mutual information formula in the main text. The grouping includes complete temporal wavepackets, so no emitted inter-time coherence has been removed. This single-emission benchmark contains one quantum in a superposition of emission times, not a simulation of a many-quantum black-hole evaporation history. The spectral obstruction, in contrast, makes no such one-excitation restriction.

\section*{S10a. A radiation-only coding benchmark}
For a diagonal input with probabilities $(x,1-x)$ and transmission probabilities $(\lambda_h,\lambda_l)$, the collected output has eigenvalues
\begin{equation}
 \bm r=(x\lambda_h,(1-x)\lambda_l,1-p_x),\qquad
 p_x=x\lambda_h+(1-x)\lambda_l.
\end{equation}
The complementary output has eigenvalues $\bm e=(x(1-\lambda_h),(1-x)(1-\lambda_l),p_x)$. These follow directly by tracing the one-excitation isometry in S6. Therefore $I_c=H(\bm r)-H(\bm e)$. For independent uses of this channel, the quantum coding theorem implies an unassisted achievable rate at least $\max(0,I_c)$\cite{Devetak}. The initial state and bath are independently prepared for each use; this does not assume independent emissions within one unmodeled black-hole history.

For the main qubit example, numerical scalar maximization over diagonal inputs gives $x=0.1558153$, $I_c=0.08607388$, $p_x=0.34607593$. Positivity does not depend on the precision of this maximization: inserting, for example, $x=0.156$ in the displayed entropy expression already gives a strictly positive rate. Full Kraus density matrices and their complements independently reproduce the expression. We quote a lower bound, not the globally regularized quantum capacity, and do not claim a finite-block code construction. Such capacity questions for generalized erasure channels have substantial precedents\cite{Filippov}.

To match the actual collected energy for this coding ensemble, the isotropic comparison uses transmission $p_x$ rather than the maximally-mixed mean $1/2$. With $\Gamma=4$, $\gamma=1$ and $\eta=1$, S9 realizes it with $\kappa=p_x/(0.8-p_x)$ and $g=\sqrt\kappa$. Its erasure channel is antidegradable and has zero unassisted capacity. Every source input still carries one degenerate excitation, and both channels completely emit it. Consequently the mean initial and collected carrier energies are matched, while the transmission dependence on the encoded state differs. The comparison does not match the full time-resolved flux or spectrum. The 147-spectrum scan uses $(\lambda_h,\lambda_l)=(1/2+s,1/2-s)$ for $0\le s\le0.29$ plus the frontier point, keeping both transmissions in the attainable range below $0.8$.

\section*{S11. Finite SYK implementation and the spectral example}
For $N=2n$ Majoranas on $n$ qubits, use
\begin{equation}
 \chi_{2j-1}=\frac{Z_1\cdots Z_{j-1}X_j}{\sqrt2},\qquad
 \chi_{2j}=\frac{Z_1\cdots Z_{j-1}Y_j}{\sqrt2}.
\end{equation}
The code constructs quartic products as monomial matrices and checks an independent dense multiplication implementation for $N=8$. The random variance is $6J^2/N^3$, and all ordered quartic products are Hermitian. The operator $2i\chi_1\chi_2$ is Hermitian with norm one. Exact diagonalization yields $E_n$ and $O_{mn}$. For $\omega_{nm}>0$, the downward and upward thermal line weights satisfy
\begin{equation}
 w_{n\to m}=p_n|O_{mn}|^2=e^{-\beta\omega_{nm}}p_m|O_{nm}|^2
 =e^{-\beta\omega_{nm}}w_{m\to n}.
\end{equation}
The continuum normalization of the main Hamiltonian gives the factor $2\pi J_\alpha(\omega)$ in Fermi's golden rule. Transition probabilities determine the initial power diagnostic but not a full phase-sensitive recovery channel. The exact source and all line weights are retained in the reproducible calculation; no phenomenological classical noise is fitted.

The infrared consistency of the chosen bath can be checked by power counting. In the conformal window of the large-$N$, $q=4$ source, the fermion two-point function has $G_\chi(\tau)\propto\operatorname{sgn}(\tau)/(J|\tau|)^{1/2}$, so $\Delta_\chi=1/4$\cite{Maldacena}. For the fixed-index, nonsinglet even operator $O=2i\chi_1\chi_2$, leading factorization gives $\langle O(\tau)O(0)\rangle\simeq4G_\chi(\tau)^2\propto|\tau|^{-1}$ and $\Delta_O=1/2$. This is not a claim about the dimension of a singlet collective bilinear. The vacuum bath has
\begin{equation}
 \langle B(\tau)B(0)\rangle=\int_0^\infty d\omega\,J_{\rm bath}(\omega)e^{-\omega|\tau|}
 \propto |\tau|^{-3}
\end{equation}
at long times because $J_{\rm bath}(\omega)\propto\omega^2$ near zero frequency. Thus $\Delta_B=3/2$. A weak boundary perturbation $h\int du\,O(u)B(u)$ has coupling dimension $1-\Delta_O-\Delta_B=-1$ and is irrelevant in this conformal power counting. A fixed number of such weak probes also does not change the leading extensive source action. This supplies a controlled infrared rationale for the probe regime. It does not establish that finite-size numerics lie in that regime, that backreaction remains negligible throughout complete evaporation, or that the specific bath has a unique four-dimensional bulk interpretation. The recovery obstruction does not require these extra assertions.

For the two-channel nonproportional example in main Fig.~2a set
\begin{align}
 J(\omega)&=\frac{0.03}{1+\omega^2}
 \begin{pmatrix}1&0.25\\0.25&0.7\end{pmatrix}=L(\omega)L(\omega)^T,\\
 V(\omega)&=\begin{pmatrix}\cos\phi&-\sin\phi\\\sin\phi&\cos\phi\end{pmatrix},\qquad
 \phi=0.8\sin\omega,\\
 a_1(\omega)&=0.25+0.38e^{-(\omega-1.4)^2/0.2},\qquad
 a_2(\omega)=0.18+\frac{0.12\omega}{1+\omega},\\
 J_R&=LV\operatorname{diag}(a_1,a_2)V^TL^T,\qquad J_E=J-J_R.
\end{align}
Both are positive semidefinite. The generalized collection eigenvalues are $a_1,a_2$ although $J_R,J_E$ need not commute. This is a spectral diagnostic example, not another gravitational model. Its above-threshold window shows failure of a sufficient no-go condition, not proof of a recoverable channel.

\section*{S12. Independent verification and reproducibility}
The code provides analytical formulas, direct amplitudes, independent population dynamics, unrestricted primal/dual recovery optimization, canonical-operator tests and raw tables. The recovery programs follow the established SDP formulation\cite{Fletcher}. For channel Kraus matrices $K_a:C\to R$, vectorize decoder matrices in output-code-major order and form
\begin{equation}
 M=k^{-2}\sum_a v_av_a^\dagger,\quad v_a=\operatorname{vec}(K_a^{\dagger}),\qquad
 \max_{J_D\succeq0,\,\Tr_CJ_D=I_R}\Tr(MJ_D).
\end{equation}
Here the code's flattening convention is equivalent to row-major vectorization of $K_a^\dagger$. Its dual is
\begin{equation}
 \min_Y\Tr Y\quad\text{subject to}\quad I_C\otimes Y\succeq M.
\end{equation}
No recovery ansatz or analytical optimum is inserted in these programs. Numerical repair of primal and dual feasibility is reported alongside raw objectives. The repaired matrices are checked to floating-point tolerance; they are not rigorous interval certificates. Two complex dual instances trigger an SCS fallback after CLARABEL's accuracy warning; the final statuses are optimal and the residuals are retained.

Supplementary Fig.~S1 presents the independent decoder and dynamical checks. Main Fig.~3e,f shows code-size scaling and sampled encodings.

The verification report has 28 groups. They include 606 independent constrained scalar optimizations of the yield frontier, 2,000 random contraction spectra, 150 random encodings per dimension pair, complete Fock operator-basis checks of the antidegrader through input occupation six, reference-entangled stability tests, exact no-cloning projector norms, conservation and positivity tests, and finite-source detailed balance. The tested code dimensions, source sizes and seeds are explicit in the tables. No disorder error bar is used to infer a thermodynamic-limit claim from three samples.

Pinned dependencies are NumPy 2.5.4, SciPy 1.18.1, Matplotlib 3.10.8, CVXPY 1.9.3, CLARABEL 0.11.1 and SCS 3.3.1; the run used Python 3.12.14. All scripts and numerical tables are collected in \texttt{code\_and\_data/}. Install its \texttt{requirements.txt} and run \texttt{python run.py} from that directory to recompute, verify and plot the results. The manuscripts can be compiled separately with \texttt{python build\_pdfs.py} from the package root. Fresh-directory verification regenerates the numerical tables before comparison with the supplied results. REVTeX support files are included with their notices; working \texttt{latex}, \texttt{dvips} and \texttt{ps2pdf} installations are required.

\section*{S13. Contribution and comparison with prior work}
Bosonic-loss antidegradability, generalized erasure, pseudomode memory and the low-energy SYK correspondence are established foundations\cite{Caruso,Filippov,Pleasance,Maldacena}. Microscopic SYK--bath studies already address entropy and information retrieval\cite{ChenQi,ChenMemory}, and quantum-graph models investigate interference-induced communication enhancement\cite{Catalano}. The appearance of memory, an information revival or a positive coding rate is therefore not itself the distinction pursued here.

The central result is the exact deterministic entanglement-fidelity frontier at fixed mean collection for the specified one-excitation channel. The proof supplies the optimal spectrum for every code dimension and yield. Its qubit point is attained by a fixed-coupling emitter. A separate min--max argument selects the best permitted isometric code for a fixed transmission map. The matrix spectral criterion translates the known loss obstruction into a condition on microscopic source--bath couplings, with a conservative stability estimate for small violations.

These statements concern different tasks. The general spectral theorem allows an interacting source and arbitrary emission multiplicity, while the sharp frontier assumes one excitation and a vacuum flag. Its average-fidelity gain permits an input-state trade-off. The coding example uses independent preparations and matches the chosen ensemble's actual output yield. Collection of the environment, source-controlled echoes and postselection are excluded throughout the radiation-only receiver task. These assumptions define the contribution precisely and prevent established assistance or loss effects from being presented as an unconditional benefit of memory.

\section*{S14. Microscopic amplitudes and the gravitational completion}
A source-to-radiation isometry for allowed initial states $|i\rangle$ can be written
\begin{equation}
 V|i\rangle=\sum_{r,e}A^{(i)}_{re}|r\rangle_R|e\rangle_{\bar R},\qquad
 \sum_{r,e}A^{(i)*}_{re}A^{(j)}_{re}=\delta_{ij}.
\end{equation}
Here $\bar R$ includes residual source degrees of freedom and inaccessible outputs. Population observables constrain expressions such as $P(r|i)=\sum_e|A^{(i)}_{re}|^2$. Recovery of an unknown input requires the complete channel, including
\begin{equation}
 \cN_R(|i\rangle\langle j|)=\sum_{r,r',e}
 A^{(i)}_{re}A^{(j)*}_{r'e}|r\rangle\langle r'|.
\end{equation}
Diagonal populations therefore constrain less information than the coherence-sensitive channel. Specifying the Hamiltonian and source operators supplies these amplitudes within a model; connecting them to a physical gravitational source additionally requires its microscopic state space, dynamics and outgoing-field dictionary.

A simple example makes this distinction exact. Take a degenerate one-excitation radiation basis $|j\rangle_R$ and an environment with degenerate zero-energy record states. The two isometries
\begin{equation}
 V_{\rm coherent}|j\rangle=|j\rangle_R|e_0\rangle_E,\qquad
 V_{\rm record}|j\rangle=|j\rangle_R|e_j\rangle_E,
 \quad \langle e_i|e_j\rangle=\delta_{ij},
\end{equation}
have identical emitted-energy and radiation-basis populations for every input. The first gives the identity radiation channel; the second gives complete dephasing in that basis. Their optimal entanglement fidelities are respectively $1$ and $1/k$, and their unassisted capacities are $\log_2 k$ and zero. The dephasing result follows because its output is separable from a reference, whose overlap with a maximally entangled state is at most $1/k$; the identity decoder attains that value. This is an illustration of what population data determine, not a choice of black-hole dynamics. The script \texttt{gravitational\_map\_check.py} verifies the two channels' isometries, populations and reference overlaps for several dimensions.

The present calculations derive amplitudes for the specified autonomous emitter and a spectral criterion for a general interacting source. A four-dimensional gravitational application would supply the amplitudes from justified black-hole microphysics while reproducing its semiclassical emission limit and dynamical energy balance. Recovery could then be tested on that physically derived map using precisely the access conditions defined here.

\section*{S15. Independent numerical formulations}
The main Methods specify the Choi convention, primal and dual recovery programs, feasibility repair, and independent population and complex-amplitude calculations. The following comparisons test these implementations. A fresh-directory run regenerates the original 16 numerical tables before comparing them with the release.
\renewcommand{\thefigure}{S\arabic{figure}}
\makeatletter\renewcommand{\fnum@figure}{Supplementary Fig.~S\arabic{figure}}\makeatother
\begin{figure}[h!]
\centering\includegraphics[width=\textwidth]{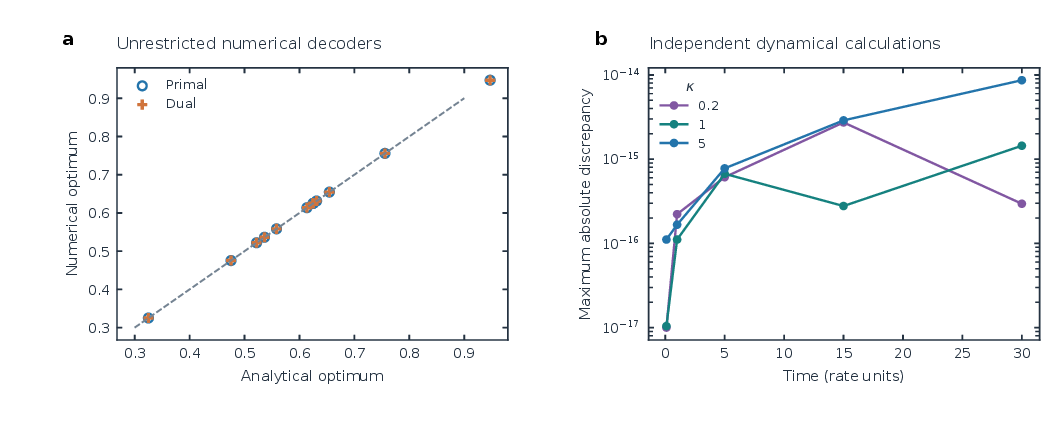}
\caption{\textbf{Independent numerical recovery and dynamics.} \textbf{a}, Twelve unrestricted recovery semidefinite programs compared with the exact analytical optimum, including complex rectangular transmission matrices and code dimensions two to four. Primal and dual optimizations are solved separately; the diagonal marks exact agreement, with maximum discrepancy $1.89\times10^{-8}$. \textbf{b}, Complex two-amplitude exponentiation followed by Gauss--Legendre integration is compared with real population-generator exponentiation. The plotted maximum includes source and memory populations, their imaginary coherence, collected probability and inaccessible probability. Three bandwidths test different dynamical regimes. The maximum absolute difference is $8.66\times10^{-15}$; a plotting floor of $10^{-17}$ displays exact numerical zeros. These checks support the implementation, while the general recovery statements are established by the preceding proofs.}\label{fig:diagnostics}
\end{figure}

\section*{S16. Reference correlations for deterministic recovery}
For a qubit, write $h=\lambda_h\ge l=\lambda_l$ and use the transmission eigenbasis. The optimal decoder reverses the successful-output isometry and prepares the weakly transmitted state $|1\rangle$ on vacuum. The composite decoded channel has Kraus operators
\begin{equation}
 K_0=\begin{pmatrix}\sqrt h&0\\0&\sqrt l\end{pmatrix},\qquad
 K_1=\sqrt{1-h}\,|1\rangle\langle0|,\qquad
 K_2=\sqrt{1-l}\,|1\rangle\langle1|.
\end{equation}
They obey $\sum_\mu K_\mu^\dagger K_\mu=I$. Acting on one half of $|\Phi_2\rangle=(|00\rangle+|11\rangle)/\sqrt2$ gives
\begin{equation}
 \rho_{AC}=\frac12\big[h|00\rangle\langle00|+|11\rangle\langle11|
 +(1-h)|01\rangle\langle01|
 +\sqrt{hl}(|00\rangle\langle11|+|11\rangle\langle00|)\big].
\end{equation}
The transverse correlations are $\langle XX\rangle=\sqrt{hl}$ and $\langle YY\rangle=-\sqrt{hl}$, while $\langle ZZ\rangle=h$. The Bell projector satisfies
\begin{equation}
 |\Phi_2\rangle\langle\Phi_2|=\tfrac14(I\otimes I+X\otimes X-Y\otimes Y+Z\otimes Z),
 \qquad F_e=\frac{1+h+2\sqrt{hl}}4.
\end{equation}
This reproduces the exact decoder formula for $k=2$. At half mean collection, the frontier gives transverse magnitude $1/\sqrt5$ and longitudinal correlation $(1+1/\sqrt5)/2$. Equal transmission gives transverse magnitude and longitudinal correlation $1/2$. The mean collected probability is the same, while the reference correlations resolve how the gain is distributed. The finite-use fidelity improvement is not an improvement for every input and is not a capacity formula.

To estimate these correlations in a controlled implementation, prepare the reference--input Bell state, apply the source-to-radiation channel and the specified deterministic decoder, and measure the three matching Pauli settings on separate repetitions. Include the vacuum fallback outcomes in every setting. Applying the identity to the measured means estimates $F_e$ without a success-conditioned renormalization. Detector loss, calibration errors and imperfect preparation alter the implemented channel and must be characterized before comparing a device with the ideal prediction. No hardware feasibility or astronomical observation is inferred here.

The script \texttt{observables.py} checks the Kraus output, Pauli estimator and exact formula for ten benchmark spectra and one hundred random ordered pairs. Predictions and residuals are saved in \texttt{operational\_observables.csv} and \texttt{observable\_checks.json}.

\begingroup\renewcommand{\bibfont}{\fontsize{8.8}{10.3}\selectfont}

\endgroup